\documentclass[journal]{IEEEtran}
\usepackage{cite}

\usepackage{graphicx}
\usepackage{xcolor}
\graphicspath{{./figure/}} 

\usepackage[fleqn]{amsmath} 
\usepackage{amssymb,bm,mathtools}
\usepackage{amsthm} 
\usepackage{algorithm,algorithmic}

\usepackage{newtxtext}

\makeatletter
\let\MYcaption\@makecaption
\makeatother

\usepackage[font=footnotesize]{subcaption}

\makeatletter
\let\@makecaption\MYcaption
\makeatother

\usepackage{subcaption}

\usepackage{url}
\usepackage{ascmac} 
\usepackage{type1cm} 
\usepackage{float}
\usepackage{pifont}
\usepackage{afterpage}
\usepackage{comment}
\usepackage{subcaption} 
\usepackage{cleveref} 
\usepackage{siunitx} 
\DeclareMathOperator*{\minimize}{minimize}
\DeclareMathOperator*{\argmin}{arg~min} 

\DeclareMathOperator{\prox}{prox}

\newcommand{\st}{\mathrm{subject~to}}

\newcommand{\abs}[1]{\left\lvert#1\right\rvert}
\newcommand{\norm}[1]{\left\lVert#1\right\rVert}

\newcommand{\paren}[1]{\left(#1\right)}
\newcommand{\sqb}[1]{\left[#1\right]}

\newcommand{\GammaO}[1]{\Gamma_{0}(#1)}

\allowdisplaybreaks[1]

\begin{document}

\title{Optimization-Based Image Restoration under Implementation Constraints in Optical Analog Circuits}

\author{Taisei Kato, Ryo Hayakawa,~\IEEEmembership{Member,~IEEE}, Soma Furusawa, Kazunori Hayashi,~\IEEEmembership{Member,~IEEE}, and Youji Iiguni% <-this % stops a space
\thanks{This work will be submitted to the IEEE for possible publication. Copyright may be transferred without notice, after which this version may no longer be accessible.}%
\thanks{This work was supported in part by JST CREST JPMJCR21C3 and JSPS KAKENHI Grant Number 25H01111.}%
\thanks{Taisei Kato and Youji Iiguni are with Graduate School of Engineering Science, The University of Osaka, 1-3 Machikaneyama, Toyonaka, Osaka, 560-8531, Japan.}%
\thanks{Ryo Hayakawa is with Institute of Engineering, Tokyo University of Agriculture and Technology, 2-24-16 Naka-cho, Koganei, Tokyo, 184-8588, Japan.}
\thanks{Soma Furusawa and Kazunori Hayashi are with the Graduate School of Informatics, Kyoto University, Kyoto, Japan.}%
}

\markboth{}%
{Optimization-Based Image Restoration under Implementation Constraints in Optical Analog Circuits}


\maketitle

\begin{abstract}
Optical analog circuits have attracted attention as promising alternatives to traditional electronic circuits for signal processing tasks due to their potential for low-latency and low-power computations. 
However, implementing iterative algorithms on such circuits presents challenges, particularly due to the difficulty of performing division operations involving dynamically changing variables and the additive noise introduced by optical amplifiers. 
In this study, we investigate the feasibility of implementing image restoration algorithms using total variation regularization on optical analog circuits. 
Specifically, we design the circuit structures for the image restoration with widely used alternating direction method of multipliers (ADMM) and primal dual splitting (PDS). 
Our design avoids division operations involving dynamic variables and incorporate the impact of additive noise introduced by optical amplifiers. 
Simulation results show that the effective denoising can be achieved in terms of peak signal to noise ratio (PSNR) and structural similarity index measure (SSIM) even when the circuit noise at the amplifiers is taken into account.
\end{abstract}


\section{Introduction}
\label{sec:introduction}
\IEEEPARstart{T}{o} realize signal processing with low latency and low power consumption, optical devices have attracted attention as alternatives to conventional electronic devices~\cite{shlezinger2021dynamic}. 
Unlike traditional circuits that use electrons as information carriers, optical analog circuits use light to carry information. 
Therefore, they are expected to perform operations such as matrix-vector multiplication with lower latency and reduced power consumption compared to their electronic counterparts~\cite{shen2017deep,wetzstein2020inference,zhou2021large,huang2022prospects,zhou2022photonic}.

However, signal processing with optical analog circuits requires consideration of several implementation constraints. 
Firstly, in iterative computations, it is challenging to implement division operations involving dynamic variables that change at each iteration. 
As a result, in the context of compressed sensing for sparse signal recovery, algorithms that avoid such operations and are tailored for optical analog circuits have been proposed~\cite{kameda2022performance,lin2022approximated}. 
Secondly, signal amplification within the circuit introduces additive noise depending on the amplification gain. 
Thus, the performance of compressed sensing algorithms have been evaluated under the effect of circuit noise due to amplification~\cite{furusawa2023numerical}. 
Simulation results have demonstrated that the performance remains comparable to the case without considering such noise.
Although compressed sensing has been the primary focus of these studies, the applicability of optical analog circuits to broader signal processing problems has yet to be thoroughly investigated. 

Image restoration is the task of reconstructing high-quality images from corrupted, degraded, or incomplete images. 
It has various applications in areas such as medical imaging, satellite imaging, and surveillance camera footage~\cite{banham1997,Kursat2012-la}. 
One of the fundamental approaches for image restoration involves solving an optimization problem using total variation (TV) regularization~\cite{rudin1992nonlinear,Chambolle2004-ej}, which promotes spatial smoothness. 
To solve such optimization problems, alternating direction method of multipliers (ADMM)~\cite{eckstein1992douglas,Boyd2011-ci} and primal-dual splitting (PDS)~\cite{condat2013primal} are commonly used in the field of image processing~\cite{Chambolle2011-yu,Komodakis2015-ze}. 
However, their implementation on optical analog circuits presents challenges, including division by dynamic variables and noise introduced by optical amplifiers. 

In this study, we investigate ADMM and PDS as candidate algorithms suitable for implementation on optical analog circuits to solve optimization problems incorporating total variation regularization for image restoration. 
Through an examination of the corresponding optical analog circuit configurations for the update equations of ADMM and PDS, we found that division by variables can be avoided if part of the computation is performed in advance using electronic circuits. 
On the other hand, since optical amplifiers are required to adjust the scale of signals within the circuit, the additional noise introduced by these amplifiers must be taken into account. 
Through computer simulations, we demonstrate that effective image denoising is still achievable even when considering the noise introduced by optical amplifiers.

Throughout this paper, we use the following notation. 
The set of all real numbers and complex numbers are denoted by $\mathbb{R}$ and $\mathbb{C}$, respectively. 
The imaginary unit is represented by $j$, i.e., $j^2 = -1$. 
Vectors (e.g., $\bm{x} \in \mathbb{R}^N$) and matrices (e.g., $\bm{A} \in \mathbb{R}^{M \times N}$) are denoted by bold lowercase and bold uppercase letters, respectively. 
The transpose of a matrix is represented as $(\cdot)^{\top}$. 
The $\ell_p$-norm of a vector $\bm{x}$ is denoted by $\|\bm{x}\|_p = \paren{\sum_{n=1}^{N} \abs{x_n}^{p}}^{\frac{1}{p}}$; for instance, the $\ell_2$-norm is $\|\bm{x}\|_2 = \sqrt{\sum_{n=1}^{N} x_n^2}$.
The set of all proper, convex, and lower semi-continuous functions from $\mathbb{R}^N$ to $(-\infty, +\infty]$ is denoted by $\Gamma_0(\mathbb{R}^N)$.
For any function $g \in \Gamma_0(\mathbb{R}^N)$ and any scalar $\gamma > 0$, the proximal operator of $\gamma g$ is defined as
\begin{align}
    \mathrm{prox}_{\gamma g}(\bm{x}) \coloneqq \argmin_{\bm{u} \in \mathbb{R}^N} \left\{ g(\bm{u}) + \frac{1}{2\gamma} \|\bm{x}-\bm{u}\|_2^2 \right\}. 
    \label{eq:prox_operator_definition} 
\end{align}

The remainder of this paper is organized as follows.
Section~\ref{sec:circuit} first explains the components and constraints of optical analog circuits.
Section~\ref{sec:proximal} describes optimization algorithms based on proximal splitting.
Section~\ref{sec:compressed_sensing} introduces related work on compressed sensing algorithms suitable for optical analog circuits. 
In Section~\ref{sec:proposed}, we describe the image restoration algorithm used in this study and then newly discuss its implementation with optical analog circuits. 
Section~\ref{sec:result} presents the restoration accuracy and restored images obtained through computer simulations. 
Finally, Section~\ref{sec:summary} concludes this paper and discusses future work.

\section{Optical Analog Circuits}\label{sec:circuit}

\subsection{Components of Optical Analog Circuits}\label{subsec:circuit}

Optical analog circuits, which are currently under development, utilize light as the carrier of information, in contrast to conventional electronic circuits that rely on electrons. 
Optical analog circuits are expected to enable operations such as matrix-vector multiplication with lower latency and reduced power consumption~\cite{shlezinger2021dynamic}.
Optical analog circuits consist of components such as signal splitters (SS), adders, subtractors, multipliers, attenuators, amplifiers, and delay elements. 
Among these, the SS, adder, and subtractor can all be realized using a common optical device known as a beam splitter (BS)~\cite{furusawa2023numerical}.

\begin{figure}[t!]
    \centering 
    \includegraphics[width=0.9\linewidth]{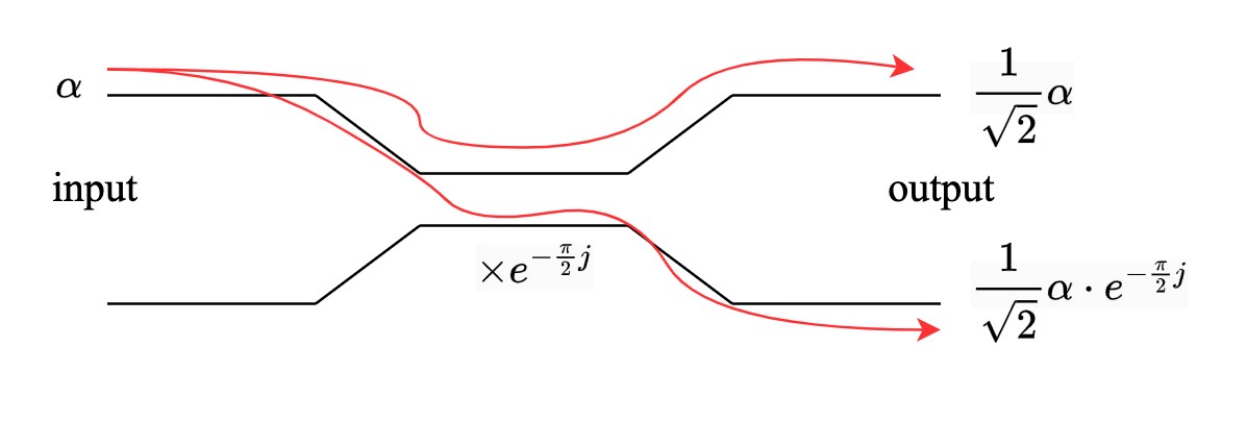} 
    \caption{Beam Splitter (BS)} 
    \label{BS} 
\end{figure}

As shown in Fig.~\ref{BS}, BS has two input ports and two output ports. 
When an arbitrary signal $\alpha \in \mathbb{C}$ is input into one of the ports, the same-side output port outputs $\frac{1}{\sqrt{2}}\alpha$, while the opposite-side port outputs $\frac{1}{\sqrt{2}}\alpha \cdot e^{-\frac{\pi}{2}j}$. 
The power of each output signal becomes half of the input signal power.

\begin{figure}[t!]
  \centering 
  \includegraphics[width=0.9\linewidth]{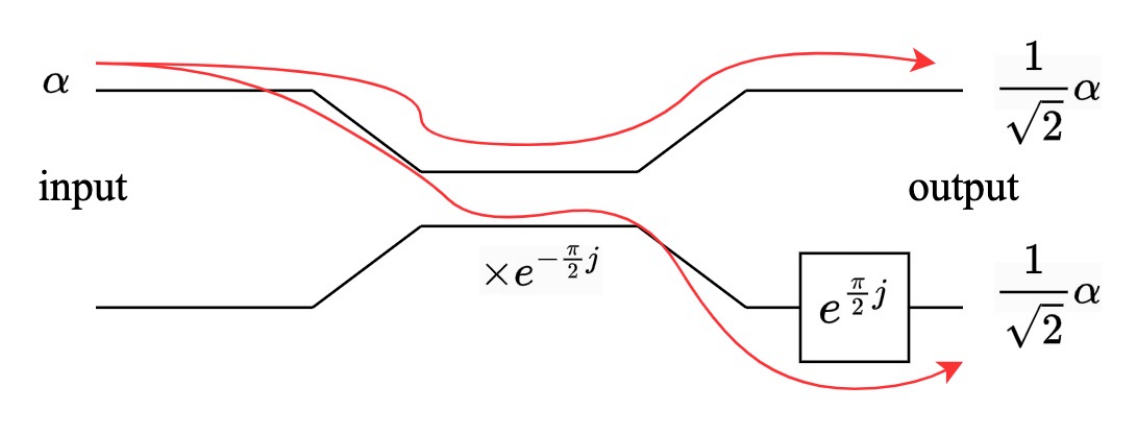} 
  \caption{Signal Splitter (SS)}
  \label{SS}
\end{figure}

SS is implemented by applying a phase shift through multiplication by $e^{\frac{\pi}{2}j}$ at the lower port of the BS, as shown in Fig.~\ref{SS}. 
As a result, the output signal power of the SS is also half of the original input power.

\begin{figure}[t!]
    \centering
    \includegraphics[width=1.0\linewidth]{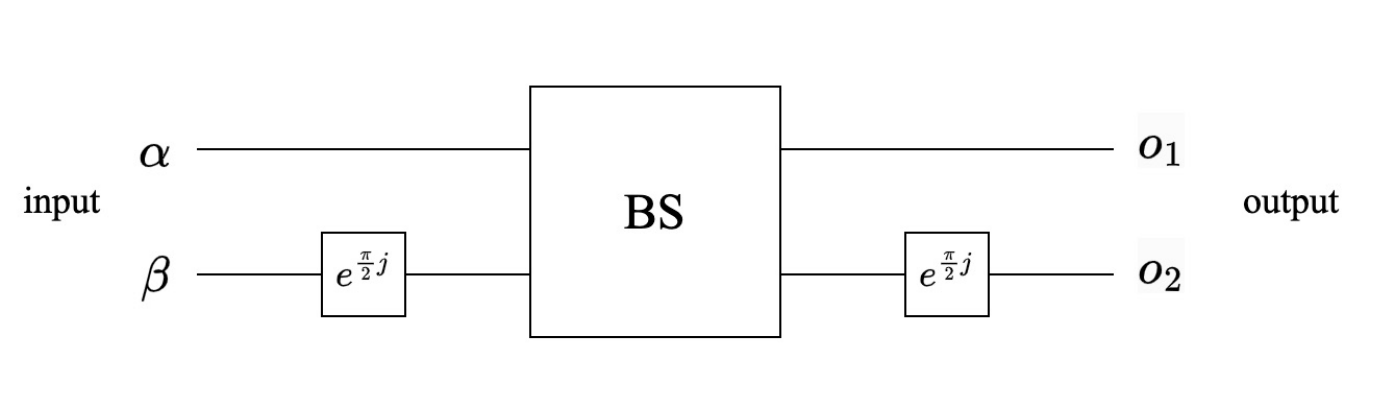}
    \caption{Optical Adder/Subtractor}
    \label{optical}
\end{figure}

Adders and subtractors are also realized using the BS and phase shifts.
As shown in Fig.~\ref{optical}, when two arbitrary signals $\alpha, \beta \in \mathbb{C}$ are input into two different input ports, the signals output from the two output ports are 
\begin{align}
    o_1 &= \frac{1}{\sqrt{2}}\alpha + \frac{1}{\sqrt{2}}\beta \cdot e^{\frac{\pi}{2}j} \cdot e^{-\frac{\pi}{2}j} \\
        &= \frac{1}{\sqrt{2}}(\alpha + \beta),
    \label{o-1}\\
    o_2 &= \frac{1}{\sqrt{2}}\alpha \cdot e^{-\frac{\pi}{2}j} \cdot e^{\frac{\pi}{2}j} + \frac{1}{\sqrt{2}}\beta \cdot e^{\frac{\pi}{2}j} \cdot e^{\frac{\pi}{2}j} \\
        &= \frac{1}{\sqrt{2}}(\alpha - \beta)
    \label{o-2}
\end{align}
Again, it is important to note that, as in the case of the SS, the output signal power is half of the original output power.

\subsection{Structural Constraints in Optical Analog Circuits}
Optical analog circuits are subject to structural constraints that differ from those in electronic circuits.
Particularly, implementing division operations in computations involving dynamic variables---such as those encountered in iterative algorithms---presents a significant challenge.
Furthermore, as described in Section~\ref{subsec:circuit}, signals in optical analog circuits are attenuated by components such as adders, subtractors, and signal splitters (SS).
One method to compensate for this attenuation is signal amplification using optical amplifiers; however, this approach inevitably introduces additional noise with each amplification.

In this section, we discuss the power of additive noise based on the noise model of optical amplifiers~\cite{furusawa2023numerical}. 
In optical fiber communications, erbium doped fiber amplifiers (EDFAs) are commonly used. 
The power spectral density of amplified spontaneous emission (ASE) noise from EDFAs is given by
\begin{align}
    G_{\rm{ASE}}=F(G-1)h\mu
    \label{ASE}
\end{align}
as in~\cite{poggiolini2012detailed}, where $F$ is the noise figure (NF) of the amplifier, $G$ is its power gain, $h$ is Planck's constant, and $\mu$ is the frequency.
Assuming the use of optical devices in a typical optical fiber communication system, we consider a wavelength of \SI{1550}{\nano\meter} (corresponding to a frequency of approximately \SI{1.94e14}{\hertz}) and a signal bandwidth of \SI{10}{\giga\hertz}.
Under this assumption, the power of ASE noise added by an optical amplifier with power gain $G$ is given by
\begin{align}
    G_{\rm{ASE}}\times \SI{10}{\giga\hertz}
    =
    (G-1) \cdot 2.56 \times10^{-9}, \label{power}
\end{align}
where the ideal NF value of $F=2$ was used. 
TABLE~\ref{table(noise)} shows the added noise power for optical amplifiers with different power gains $G$.
\begin{table}[tb]
  \caption{Estimated added noise power for amplifiers with different power gains~\cite{furusawa2023numerical}}
  \label{table(noise)}
  \centering
  \begin{tabular}{cc}
    \hline
    Power Gain $G$  & Added Noise Power  \\
    \hline \hline
     8  & $1.79\times10^{-8}$ \\
    16  & $3.84\times10^{-8}$ \\
    32 & $7.94\times10^{-8}$ \\
    64  &  $1.61\times10^{-7}$ \\
    128  &  $3.25\times10^{-7}$ \\
    256  &  $6.53\times10^{-7}$ \\
    \hline
  \end{tabular}
\end{table}
This indicates that when a signal is amplified by an optical amplifier by a factor of its power gain $G$, the amount of noise power shown in TABLE~\ref{table(noise)} is added.

\section{Proximal Splitting Algorithms} \label{sec:proximal}
In this section, we provide an overview of representative algorithms for solving optimization problems whose objective function is a sum of a differentiable term and a non-differentiable term whose proximal operator is easy to compute.

\subsection{Proximal Gradient Method}
We consider the problem of minimizing the sum of two functions given by 
\begin{align}
    \minimize_{\bm{x} \in \mathbb{R}^{N}} \ f(\bm{x}) + g(\bm{x}). 
    \label{prob:prox_grad}
\end{align}
Here, we assume that $f:\mathbb{R}^{N} \to \mathbb{R}$ is a differentiable convex function and its gradient $\nabla f$ is Lipschitz continuous. 
Moreover, $g: \mathbb{R}^{N} \to (-\infty, +\infty]$ is a convex function such that $g \in \GammaO{\mathbb{R}^{N}}$, where $\GammaO{\mathbb{R}^{N}}$ denotes the set of all proper, lower semi-continuous convex functions on $\mathbb{R}^{N}$. 
For this type of problem, the proximal gradient method~\cite{Combettes2005-mf,daubechies2004iterative} has been proposed. 
The update equations of PGM consist of a gradient descent step and a proximal operator step.

\subsection{ADMM}

ADMM~\cite{eckstein1992douglas,Boyd2011-ci} is an optimization algorithm for solving optimization problems of the form
\begin{align}
    \minimize_{\bm{x} \in \mathbb{R}^{N},\ \bm{z} \in \mathbb{R}^{L}}
    \ f(\bm{x}) + g(\bm{z})  \;\;
    \st  \;\;  \bm{z}=\bm{G}\bm{x},
    \label{ADMM}
\end{align}
where $f\in\GammaO{\mathbb{R}^{N}}$, $g\in\GammaO{\mathbb{R}^{L}}$, and $\bm{G}\in\mathbb{R}^{L\times N}$.
For initial values $\bm{z}_{0}, \bm{v}_{0}\in\mathbb{R}^{L}$ and the parameter $\gamma>0$, the update equations of ADMM are described in Algorithm~\ref{alg:ADMM(2.10)}. 

\begin{algorithm}[t!]
    \caption{ADMM for~\eqref{ADMM}}
    \label{alg:ADMM(2.10)}
    \begin{algorithmic}[1]
        \REQUIRE Initial values $\bm{z}_{0}, \bm{v}_{0}$, step size $\gamma>0$
        \STATE $t \leftarrow 0$
        \WHILE{stopping criterion is not met}
            \STATE $\bm{x}_{t+1}=\underset{\bm{x}\in\mathbb{R}^{N}}\argmin\;\left\{f(\bm{x})+\frac{1}{2\gamma}\|\bm{z}_t-\bm{Gx}-\bm{v}_t\|^2_2\right\}$
            \STATE $\bm{z}_{t+1}=\prox_{\gamma g}(\bm{G}\bm{x}_{t+1}+\bm{v}_{t})$
            \STATE $\bm{v}_{t+1}=\bm{v}_{t} + \bm{G}\bm{x}_{t+1} - \bm{z}_{t+1}$
            \STATE $t\leftarrow t+1$
        \ENDWHILE
        \ENSURE $\bm{x}_{t}$
    \end{algorithmic}
\end{algorithm}

\subsection{PDS}

PDS~\cite{condat2013primal,Vu2013-bi} is an algorithm for efficiently solving problems involving the sum of multiple convex functions, particularly those with terms including linear operators. 
PDS can solve optimization problems of the form
\begin{align}
    \minimize_{\bm{x} \in \mathbb{R}^{N}}
    \ f(\bm{x}) + g(\bm{x}) + h(\bm{Gx}), 
    \label{PDS}
\end{align}
where $f,g\in\GammaO{\mathbb{R}^{N}}$, $h\in\GammaO{\mathbb{R}^{L}}$, and $\bm{G}\in\mathbb{R}^{L\times N}$.
Furthermore, it is assumed that $f$ is differentiable and its gradient $\nabla f$ is $\beta$-Lipschitz continuous ($\beta > 0$). 
For any initial values $\bm{x}_{0}\in\mathbb{R}^{N}, \bm{v}_{0}\in\mathbb{R}^{L}$ and parameters $\gamma_1,\gamma_2>0$, the update equations of PDS are described in Algorithm~\ref{alg:PDS(0)}. 
\begin{algorithm}[t!]
    \caption{PDS for~\eqref{PDS}}
    \label{alg:PDS(0)}
    \begin{algorithmic}[1]
        \REQUIRE Initial values $\bm{x}_{0}, \bm{v}_{0}$; step sizes $\gamma_1, \gamma_2 > 0$
        \STATE $t \leftarrow 0$
        \WHILE{stopping criterion is not met}
            \STATE $\bm{x}_{t+1}=\prox_{\gamma_{1} g}(\bm{x}_{t}-\gamma_{1}(\nabla f(\bm{x}_{t}) + \bm{G}^{\top}\bm{v}_{t}))$
            \STATE $\bm{z}_{t+1} = \bm{v}_{t}+\gamma_{2}\bm{G}(2\bm{x}_{t+1}-\bm{x}_{t})$
            \STATE $\bm{v}_{t+1}=\prox_{\gamma_{2} h^{*}}(\bm{z}_{t+1})$
            \STATE $t\leftarrow t+1$
        \ENDWHILE
        \ENSURE $\bm{x}_{t}$
    \end{algorithmic}
\end{algorithm}
Here, $h^{*}$ is the convex conjugate of the function $h$, and its proximal operator can be computed using the proximal operator of $h$ as
\begin{align}
    {\prox}_{\gamma h^{*}}(\bm{u})
    =
    \bm{u}-\gamma\prox_{\frac{1}{\gamma}h}\left(\frac{1}{\gamma}\bm{u}\right). 
    \label{prox(2)}
\end{align}

\section{Compressed Sensing with Optical Analog Circuits} \label{sec:compressed_sensing}

This section first describes the application of the optimization algorithms discussed in Section~\ref{sec:proximal} to compressed sensing problems. 
Then, we explain prior work~\cite{kameda2022performance,lin2022approximated} on compressed sensing algorithms designed for implementation in optical analog circuits. 
Although optical analog circuits can handle complex-valued signals, we consider real-valued signals hereafter for simplicity. 

\subsection{Problem Setup of Compressed Sensing and $\ell_1$-$\ell_2$ Reconstruction}

Compressed sensing~\cite{Candes2005-ut,donoho2006compressed} is a framework for reconstructing sparse signals (i.e., signals containing many zero components) from a small number of observations. 
Since various signals in engineering possess sparsity, compressed sensing has many applications such as image processing, wireless communications, and control engineering~\cite{Lustig2007-vj,Hayashi2013-nq,Choi2017-dk,Nagahara2020-sz}.

In compressed sensing, we consider a known measurement vector $\bm{y} \in \mathbb{R}^{M}$ which is obtained from an unknown sparse vector $\bm{x}^* \in \mathbb{R}^{N}$ through a linear process. The relationship is given by 
\begin{equation}
    \bm{y} = \bm{A}\bm{x}^* + \bm{e},
    \label{model:cs} 
\end{equation}
where $\bm{A} \in \mathbb{R}^{M \times N}$ is a known sensing matrix and $\bm{e} \in \mathbb{R}^{M}$ represents observation noise. 
Typically, the number of measurements $M$ is smaller than the dimension of the signal $N$ (i.e., $M < N$). 

A widely used method for estimating the sparse unknown vector $\bm{x}^{*}$ from the noisy observation vector $\bm{y}$ in \eqref{model:cs} is $\ell_1$-$\ell_2$ reconstruction.
This is formulated by the following optimization problem
\begin{align}
    \minimize_{\bm{x}\in\mathbb{R}^{N}} \ \frac{1}{2}\norm{\bm{Ax-y}}^{2}_{2}+\lambda\norm{\bm{x}}_{1}. \label{opt:cs}
\end{align}
The first term in the objective function of~\eqref{opt:cs} is the data fidelity term, and the second term is the $\ell_1$ regularization term that promotes the sparsity of $\bm{x}$. 
$\lambda$ ($>0$) is a regularization parameter that controls the balance between the data fidelity and the sparsity of the solution.

\subsection{FISTA with a Fixed Acceleration Parameter}
The optimization problem in~\eqref{opt:cs} can be solved efficiently by several algorithms based on the proximal gradient method, such as iterative shrinkage thresholding algorithm (ISTA)~\cite{daubechies2004iterative} and its accelerated version, fast iterative shrinkage thresholding algorithm (FISTA)~\cite{beck2009fast}.
However, in optical analog circuits, division by a dynamic variable is difficult, and thus FISTA using a conventional acceleration parameter is considered challenging to implement.
Therefore, constant inertial FISTA (CIFISTA)~\cite{kameda2022performance} has been proposed as an algorithm that fixes the acceleration parameter to a constant and employs a pseudo-acceleration scheme.
Simulation results have shown that when the coefficient value is set appropriately, CIFISTA exhibits convergence property almost equivalent to FISTA.

\subsection{ADMM and Approximated ADMM}

Since ADMM for the optimization problem \eqref{opt:cs} in compressed sensing does not involve division by dynamic values, the implementation of ADMM using optical analog circuits has been investigated in~\cite{lin2022approximated,furusawa2023numerical}.
However, ADMM for~\eqref{opt:cs} requires a matrix inversion, which might be prohibitive for large scale problems.
To address this,~\cite{lin2022approximated} has also proposed an approximated ADMM that avoids matrix inversion by using a different formulation of the ADMM algorithm.
Simulation results in~\cite{lin2022approximated} demonstrate that the approximated ADMM can achieve convergence property nearly equivalent to the conventional ADMM algorithm without the matrix inversion, though the optical analog circuit configuration becomes more complex.

\section{Image Restoration Algorithms Suitable for Implementation in Optical Analog Circuits} \label{sec:proposed}

In this section, we first describe two image restoration algorithms considered in this study. 
We then discuss the their implementation under the structual constraints and noise characteristics of optical analog circuits.

\subsection{Image Restoration Using Total Variation Regularization}

\subsubsection{Image Restoration}

Image restoration is the process of estimating an original, undegraded image from observed data that includes degradations such as camera defocus, motion blur, or missing information.
Here, we represent the original image as a vector $\bm{x}^* \in \mathbb{R}^{N}$, where $N$ is the total number of pixels. 
This vector is typically obtained by stacking the pixel values of a two-dimensional image of size $N_{1} \times N_{2}$ into a column vector, i.e., $N = N_{1}N_{2}$.
In this paper, we consider the case where the observed image $\bm{y} \in \mathbb{R}^{M}$ is written as  
\begin{align}
    \bm{y} = \bm{A}\bm{x}^* + \bm{e}.
    \label{model:image_restoration} 
\end{align}
Here, $\bm{A} \in \mathbb{R}^{M \times N}$ is a matrix describing the degradation process, such as blur or missing data, and $\bm{e} \in \mathbb{R}^{M}$ is observation noise.

\subsubsection{Optimization Problem with Total Variation Regularization}

In compressed sensing, $\ell_1$-$\ell_2$ reconstruction in~\eqref{opt:cs} is commonly used to exploit the sparsity of $\bm{x}^{\ast}$. 
However, in image restoration problems, $\bm{x}^{\ast}$ itself is not sparse in general, and thus it is necessary to consider an alternative regularization term.
A fundamental optimization formulation for image restoration is the TV-regularized problem given by
\begin{align}
    \minimize_{\bm{x}\in\mathbb{R}^{N}}
    \ 
        \frac{1}{2} \norm{\bm{Ax}-\bm{y}}^{2}_{2}
        +  \lambda {\|\bm{Dx}\|}_{1,2}. 
    \label{model(TV)}
\end{align}

Total variation, denoted by $\|\bm{Dx}\|_{1,2}$, is a commonly used measure of image smoothness and is defined as the sum of differences between adjacent pixels. 
In this paper, it is formulated as
\begin{align}
    \|\bm{Dx}\|_{1,2}=
    \sum_{i=1}^{N}\sqrt{d_{v,i}^2+d_{h,i}^2}, 
    \label{TV(u)}
\end{align}
where $d_{v,i}$ and $d_{h,i}$ represent the vertical and horizontal differences at the $i$-th pixel, respectively.
By letting $\bm{D}_{v}, \bm{D}_{h} \in \mathbb{R}^{N\times N}$ be the matrices that compute the vertical and horizontal differences of adjacent pixels, respectively, we define the matrix $\bm{D}$ by stacking these vertically as $\bm{D} \coloneqq \sqb{\bm{D}_{v}^{\top}\ \bm{D}_{h}^{\top}}^{\top} \in \mathbb{R}^{2N\times N}$.
Specifically, $\bm{D}_{v}$ and $\bm{D}_{h}$ are defined as circulant matrices with the following first rows
\begin{equation}
    v_{1,j} = 
    \begin{cases} 
        -1 & \text{if } j=1 \\
        1  & \text{if } j=N \\
        0  & \text{otherwise}
    \end{cases}
    ,
    \label{eq:v1j_definition}
\end{equation}
\begin{equation}
    h_{1,j} = 
    \begin{cases} 
        -1 & \text{if } j=1 \\
        1  & \text{if } j = N-N_{1}+1 \\
        0  & \text{otherwise}
    \end{cases}
    ,
    \label{eq:h1j_definition}
\end{equation}
where the $(i,j)$-th entries ($i,j=1,2,\ldots,N$) of $\bm{D}_{v}$ and $\bm{D}_{h}$ are denoted as $v_{i,j}$ and $h_{i,j}$, respectively. 
The mixed $\ell_{1,2}$-norm, defined by $\|\bm{z}\|_{1,2} \coloneqq \sum_{\mathfrak{g}\in\mathfrak{G}} \|\bm{z}_{\mathfrak{g}}\|_2$, is often used for promoting group sparsity in signal recovery. 
Here, $\bm{z}_{\mathfrak{g}}$ represents each group vector when the elements of $\bm{z}$ are divided into non-overlapping groups.
In our case, the vertical and horizontal differences at each pixel are treated as one group, resulting in $N$ groups in total.

Natural images generally exhibit strong correlations between adjacent pixels, resulting in small differences between neighboring pixels except at edges, where the differences become large. 
Therefore, when the vertical and horizontal adjacent pixel differences at each pixel are grouped together, $\bm{D}\bm{x}^{\ast}$ tends to exhibit group sparsity. 
This property justifies the use of total variation as an effective regularization for image restoration.

\subsubsection{ADMM for the Optimization Problem with Total Variation Regularization}

The TV-regularized optimization problem in~\eqref{model(TV)} can be expressed as the form in~\eqref{ADMM} by setting $f(\bm{x})=\frac{1}{2}\|\bm{Ax}-\bm{y}\|_2^2$, $g(\bm{z})=\lambda\|\bm{z}\|_{1,2}$, and $\bm{G}=\bm{D}$.
We can thus obtain the ADMM-based algorithm for the problem in~\eqref{model(TV)} as 
\begin{align}
    \bm{x}_{t+1} 
    &= 
    \underset{\bm{x}\in\mathbb{R}^{N}}{\arg\min}\;
    \left\{\frac{1}{2}\|\bm{Ax}-\bm{y}\|_2^2 + \frac{1}{2\gamma}\|\bm{z}_t - \bm{Dx} - \bm{v}_t\|^2_2 \right\}, \label{x(ADMM)} \\
    \bm{z}_{t+1} 
    &=
    \prox_{\gamma\lambda\|\cdot\|_{1,2}}(\bm{D}\bm{x}_{t+1} + \bm{v}_{t}), \\
    \bm{v}_{t+1} 
    &= 
    \bm{v}_{t} + \bm{D}\bm{x}_{t+1} - \bm{z}_{t+1}
\end{align}
from Algorithm~\ref{alg:ADMM(2.10)}. 
The update of $\bm{x}_{t}$ in~\eqref{x(ADMM)} is a minimization of a quadratic function with respect to $\bm{x}$, and its solution can be obtained by solving the linear system derived from setting its gradient to zero.
Hence, we can update $\bm{x}_{t}$ as 
\begin{align}
    \bm{x}_{t+1}
    =
    \paren{\bm{A}^{\top}\bm{A}
    +\frac{1}{\gamma}\bm{D}^{\top}\bm{D}}^{-1}
    \paren{\bm{A}^{\top}\bm{y}
    +\frac{1}{\gamma}\bm{D}^{\top}(\bm{z}_t-\bm{v}_t)}. 
\end{align}
Consequently, the ADMM algorithm for the optimization problem with total variation regularization is presented as Algorithm~\ref{alg:ADMM_TV}.
\begin{algorithm}[t!]
    \caption{ADMM for problem in~\eqref{model(TV)}}
    \label{alg:ADMM_TV}
    \begin{algorithmic}[1]
        \REQUIRE Initial values $\bm{z}_{0}, \bm{v}_{0}$; parameter $\gamma, \lambda >0 $
        \WHILE{stopping criterion is not met}
        {\STATE{$\bm{x}_{t+1}=(\bm{A}^{\top}\bm{A}+\frac{1}{\gamma}\bm{D}^{\top}\bm{D})^{-1}(\bm{A}^{\top}\bm{y}+\frac{1}{\gamma}\bm{D}^{\top}(\bm{z}_t-\bm{v}_t))$}}
        {\STATE{$\bm{z}_{t+1}={\prox}_{\gamma\lambda\|\cdot\|_{1,2}}(\bm{Dx}_{t+1}+\bm{v}_t)$}}
        {\STATE{$\bm{v}_{t+1}=\bm{v}_t+\bm{D}\bm{x}_{t+1}-\bm{z}_{t+1}$}}
        {\STATE{$t\leftarrow t+1$}}
        \ENDWHILE
        \ENSURE $\bm{x}_{t}$
    \end{algorithmic}
\end{algorithm}
The proximal operator of the mixed $\ell_{1,2}$-norm~\cite{yuan2006model} is a group-wise scaled soft-thresholding function as 
\begin{align}
    [{\rm{prox}}_{\gamma\lambda\|\cdot\|_{1,2}}(\bm{z})]_{\mathfrak{g}}
    =
    \max \left\{ 1-\frac{\gamma\lambda}{\|\bm{z}_{\mathfrak{g}}\|_{2}}, 0 \right\} \bm{z}_{\mathfrak{g}}, 
    \label{prox_l12}
\end{align}
where $[{\rm{prox}}_{\gamma\lambda\|\cdot\|_{1,2}}(\bm{z})]_{\mathfrak{g}}$ is the subvector of ${\rm{prox}}_{\gamma\lambda\|\cdot\|_{1,2}}(\bm{z})$ corresponding to group $\mathfrak{g}$.

\subsubsection{PDS for the Optimization Problem with Total Variation Regularization}

The TV-regularized optimization problem in~\eqref{model(TV)} can also be solved via PDS.
If we set $f(\bm{x})=\frac{1}{2}\|\bm{Ax}-\bm{y}\|_2^2$, $g(\bm{x})=0$, and $h(\bm{G}\bm{x})=\lambda\|\bm{D}\bm{x}\|_{1,2}$ ($\bm{G}=\bm{D}$) in the optimization problem in~\eqref{PDS}, we obtain the problem in~\eqref{model(TV)}.
Since the gradient $\nabla f(\bm{x})$ is given by
\begin{align}
    \nabla f(\bm{x})
    &=
    \bm{A}^{\top}\bm{A}\bm{x} - \bm{A}^{\top}\bm{y}
\end{align}
and the proximal operator of $g(\bm{x})$ becomes the identity mapping, the update equation for $\bm{x}_{t}$ is simplified as 
\begin{align}
    \bm{x}_{t+1} = \bm{x}_{t}-\gamma_{1}(\bm{A}^{\top}\bm{Ax}_t - \bm{A}^{\top}\bm{y} + \bm{D}^{\top}\bm{v}_{t}) \label{eq:pds_tv_x_deriv_final}
\end{align}

The update for the dual variable $\bm{v}_{t+1}$ is given by $\mathrm{prox}_{\gamma_{2} h^{*}}(\bm{z}_{t+1})$.
Using the definition of the proximal operator of a convex conjugate function as provided in~\eqref{prox(2)}, we can write 
\begin{align}
    \bm{v}_{t+1}
    &= \mathrm{prox}_{\gamma_{2} h^{*}}(\bm{z}_{t+1}) \\ 
    &= \bm{z}_{t+1} - \gamma_{2} \mathrm{prox}_{\frac{\lambda}{\gamma_2}\|\cdot\|_{1,2}}\paren{\frac{1}{\gamma_{2}}\bm{z}_{t+1}}.
    \label{eq:v_update_intermediate_concise}
\end{align}
The second term in~\eqref{eq:v_update_intermediate_concise} can be simplified by considering its application to each group $\mathfrak{g}$.
Applying the proximal operator for the $\ell_{1,2}$-norm in~\eqref{prox_l12} to the term, we have 
\begin{align}
    &\left[ \gamma_{2} \mathrm{prox}_{\frac{\lambda}{\gamma_2}\|\cdot\|_{1,2}} 
    \paren{\frac{1}{\gamma_{2}}\bm{z}_{t+1}} \right]_{\mathfrak{g}} \nonumber \\
    &= \gamma_2 \max\left(1 - 
    \frac{\lambda/\gamma_2}{\| (1/\gamma_2)[\bm{z}_{t+1}]_{\mathfrak{g}} \|_2}
    ,0\right) 
    \frac{1}{\gamma_2} [\bm{z}_{t+1}]_{\mathfrak{g}} \label{eq:step_before_norm_prop_concise} \\
    &= \max\left(1 - \frac{\lambda}{\| [\bm{z}_{t+1}]_{\mathfrak{g}} \|_2} ,0\right) [\bm{z}_{t+1}]_{\mathfrak{g}} \quad \label{eq:step_after_norm_prop_concise} \\
    &= \left[ \mathrm{prox}_{\lambda\|\cdot\|_{1,2}}
    (\bm{z}_{t+1}) \right]_{\mathfrak{g}}. \label{eq:simplified_group_term_concise}
\end{align}
This simplification holds for every group $\mathfrak{g}$. 
Therefore, substituting this result back into~\eqref{eq:v_update_intermediate_concise}, the update for $\bm{v}_{t+1}$ becomes
\begin{align}
    \bm{v}_{t+1}
    &= \bm{z}_{t+1} - \mathrm{prox}_{\lambda\|\cdot\|_{1,2}}(\bm{z}_{t+1}).
    \label{eq:v_update_final_form_PDS_concise}
\end{align}

As a result, the PDS-based algorithm for the TV-regularized optimization problem in~\eqref{model(TV)} is given as in Algorithm~\ref{alg:PDS_TV}.
\begin{algorithm}[t!]
    \caption{PDS for problem in~\eqref{model(TV)}}
    \label{alg:PDS_TV}
    \begin{algorithmic}[1]
        \REQUIRE Initial values $\bm{x}_{0}, \bm{v}_{0}$, parameter $\gamma_1, \gamma_2, \lambda > 0$
        \WHILE{stopping criterion is not met}
        {\STATE{$\bm{x}_{t+1}=\bm{x}_{t}-\gamma_{1}(\bm{A}^{\top}(\bm{Ax}_t - \bm{y})+\bm{D}^{\top}\bm{v}_{t})$}}
        {\STATE{$\bm{z}_{t+1}=\bm{v}_{t}+\gamma_{2}\bm{D}(2\bm{x}_{t+1}-\bm{x}_{t})$}}
        {\STATE{$\bm{v}_{t+1}=\bm{z}_{t+1}-\prox_{\lambda\|\cdot\|_{1,2}}{(\bm{z}_{t+1}})$}}
        {\STATE{$t\leftarrow t+1$}}
        \ENDWHILE
        \ENSURE $\bm{x}_{t}$
    \end{algorithmic}
\end{algorithm}
It should be noted that, while the ADMM update equations involve a matrix inversion, the PDS update equations do not.

\subsection{Image Restoration in Optical Analog Circuits}

Optical analog circuits face two challenges: division by dynamic values and circuit noise introduced during signal amplification.
In this study, we evaluate image restoration algorithms based on ADMM and PDS while taking these two issues into account. 
Although ADMM involves matrix inversion, the inverse matrix remains constant across iterations and can be precomputed, thus avoiding per-iteration division.
In this case, ADMM in Algorithm~\ref{alg:ADMM_TV} and PDS in Algorithm~\ref{alg:PDS_TV} for TV-based image restoration do not require division by dynamic values. 
Accordingly, in this work, we focus solely on the impact of circuit noise due to amplification this time.
To evaluate the impact of circuit noise, we here assume that the calculation of proximal operators can be performed accurately in electronic circuits.

Due to multiple SSs, adders, and subtractors in the circuit, optical signals attenuate as they pass through the components.
Furthermore, since proximal operators are generally nonlinear, the input signal power must be maintained within an appropriate range.
Therefore, at each iteration of the algorithm, the power scale of signals should be adjusted by inserting optical attenuators and amplifiers appropriately.
This amplification process introduces circuit noise, making it necessary to evaluate the impact of circuit noise on image restoration accuracy.
As shown in TABLE~\ref{table(noise)}, it can be seen that the standard deviation of the added noise increases with larger amplification factors.
In the following, we denote the circuit noise introduced by an amplifier with gain $G$ at the $t$-th iteration as $\bm{n}_{t}^{G}$.

\subsubsection{Analog Circuit Configuration for ADMM-based Algorithm}

\begin{figure*}[t!]
    \centering
    \includegraphics[width=0.87\linewidth]{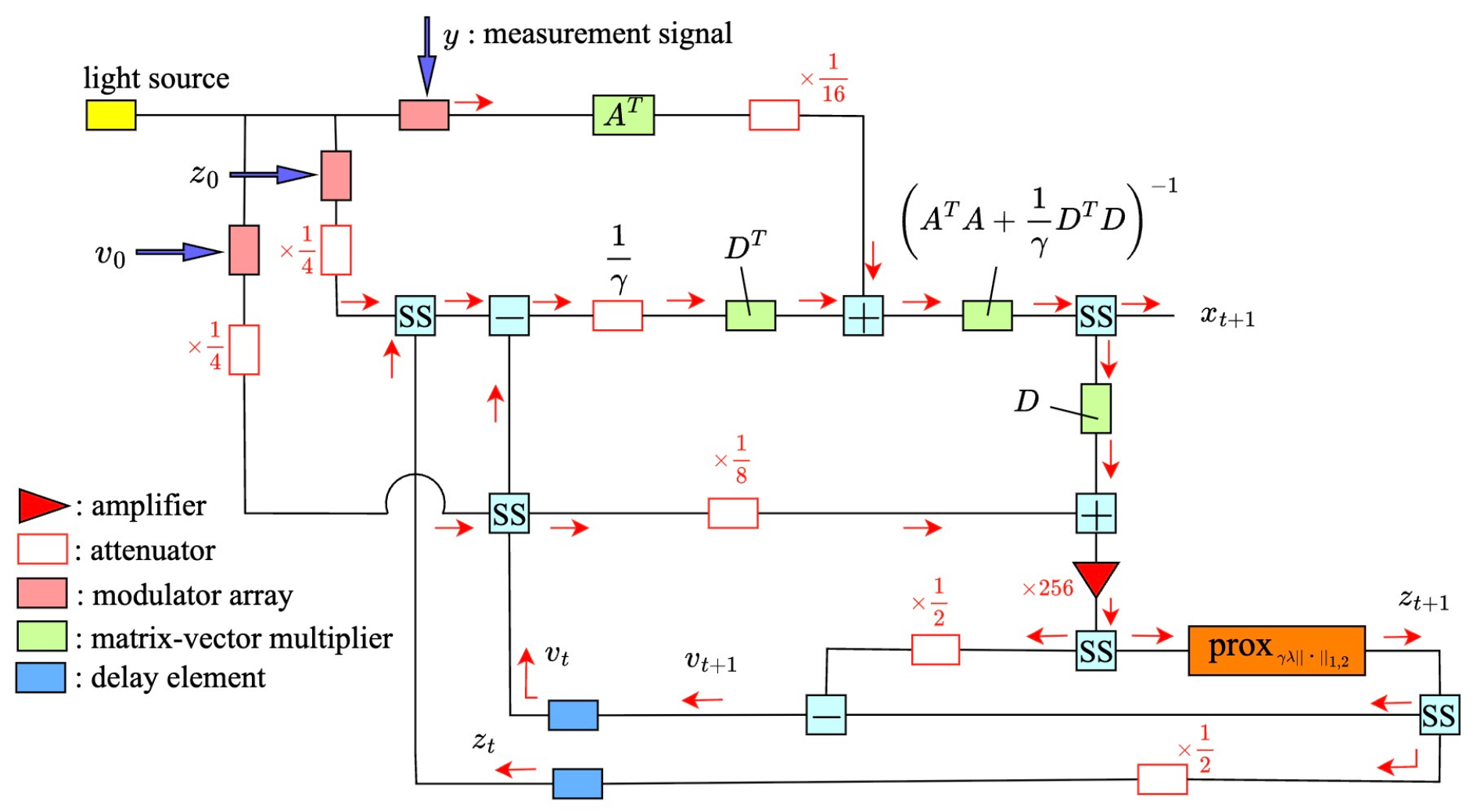}
    \caption{Optical analog circuit configuration for ADMM}
    \label{circuit-ADMM}
\end{figure*}

Fig.~\ref{circuit-ADMM} shows the optical analog circuit configuration for ADMM in Algorithm~\ref{alg:ADMM_TV} based on the optical analog circuit configuration for ADMM-based compressed sensing~\cite{furusawa2023numerical}.
As can be seen from Fig.~\ref{circuit-ADMM}, an optical amplifier with gain $G=256$ is required to compensate for signal attenuation in the circuit.
Since the amplification is located just before the proximal operator, the signal after passing through the optical amplifier becomes $\bm{Dx}_t+\bm{v}_t+\bm{n}^{256}_t$, where ${\bm{n}^{256}_t}$ is the additive noise from the optical amplifier. 
Therefore, the ADMM-based image restoration algorithm considering the circuit noise can be expressed as in Algorithm~\ref{alg:ADMM_TV(optical)}.
In addition, the update equation for $\bm{x}_{t}$ in Algorithm~\ref{alg:ADMM_TV(optical)} includes multiplication by $\frac{1}{\gamma}$. 
If $\gamma$ is less than 1, then $\frac{1}{\gamma}$ is greater than 1. 
In that case, we need to consider the cuircuit noise at the amplifier.
\begin{algorithm}[t!]
    \caption{ADMM for~\eqref{model(TV)} considering circuit noise}
    \label{alg:ADMM_TV(optical)}
    \begin{algorithmic}[1]
        \REQUIRE Initial values $\bm{z}_{0}, \bm{v}_{0}$, parameter $\gamma, \lambda >0$
        \WHILE{stopping criterion is not met}
        {\STATE{$\bm{x}_{t+1}=(\bm{A}^{\top}\bm{A}+\frac{1}{\gamma}\bm{D}^{\top}\bm{D})^{-1}(\bm{A}^{\top}\bm{y}+\frac{1}{\gamma}\bm{D}^{\top}(\bm{z}_t-\bm{v}_t))$}}
        {\STATE{$\bm{z}_{t+1}={\prox}_{\gamma\lambda\|\cdot\|_{1,2}}(\bm{Dx}_t+\bm{v}_t+\bm{n}^{256}_t)$}}
        {\STATE{$\bm{v}_{t+1}=\bm{v}_t+\bm{Dx}_{t+1}+\bm{n}^{256}_t-\bm{z}_{t+1}$}} 
        {\STATE{$t\leftarrow t+1$}}
        \ENDWHILE
        \ENSURE $\bm{x}_{t}$
    \end{algorithmic}
\end{algorithm}

\subsubsection{Analog Circuit Configuration for PDS-based Algorithm}

\begin{figure*}[t!]
    \centering
    \includegraphics[width=0.87\linewidth]{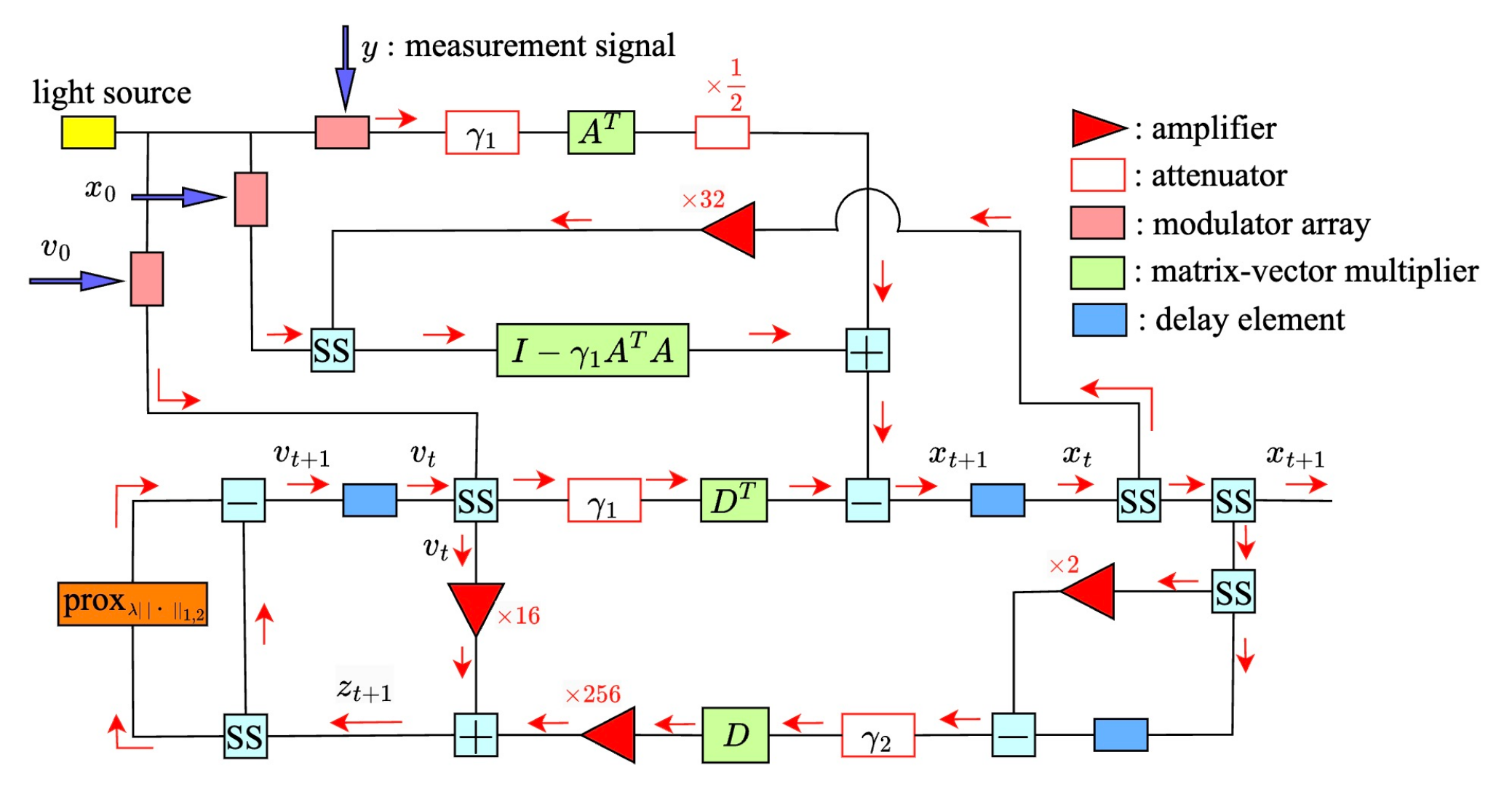}
    \caption{Optical analog circuit configuration for PDS}
    \label{circuit-PDS}
\end{figure*}

Fig.~\ref{circuit-PDS} shows the optical analog circuit configuration for PDS in Algorithm~\ref{alg:PDS_TV}.
As shown in Fig.~\ref{circuit-PDS}, PDS requires optical amplifiers with different gains. 
For each optical amplifier, additive noise vectors such as $\bm{n}^{256}_t, \bm{n}^{32}_t, \bm{n}^{16}_t, \bm{n}^{2}_t$ are introduced, depending on the amplification gain. 
The PDS-based image restoration algorithm considering the circuit noise can be written as in Algorithm~\ref{alg:PDS_TV(optical)}.
In addition, note that the update equations for $\bm{x}_{t}$ and $\bm{z}_{t}$ in Algorithm~\ref{alg:PDS_TV(optical)} include multiplication by $\gamma_{1}$ and $\gamma_{2}$, respectively. 
If the respective $\gamma$ values are greater than 1, the circuit noise at the amplifier should be taken into account.
\begin{algorithm}[t!]
    \caption{PDS for~\eqref{model(TV)} considering circuit noise}
    \label{alg:PDS_TV(optical)}
    \begin{algorithmic}[1]
        \REQUIRE Initial values $\bm{x}_{0}, \bm{v}_{0}$, parameters $\gamma_1, \gamma_2, \lambda >0$
        \WHILE{stopping criterion is not met}
            {\STATE{$\bm{x}_{t+1}=(\bm{I}-\gamma_{1}\bm{A}^{\top}\bm{A})(\bm{x}_{t}+\bm{n}^{32}_{t})+\gamma_1 \bm{A}^{\top}\bm{y} - \gamma_1 \bm{D}^{\top}\bm{v}_{t}$}}
            {\STATE{$\bm{z}_{t+1}=\bm{v}_{t}+\bm{n}^{16}_{t}+\gamma_{2}\bm{D}(2\bm{x}_{t+1}+\bm{n}^{2}_{t}-\bm{x}_{t})+\bm{n}^{256}_{t}$}}
            {\STATE{$\bm{v}_{t+1}=\bm{z}_{t+1}-\prox_{\lambda\|\cdot\|_{1,2}}{(\bm{z}_{t+1}})$}}
            {\STATE{$t\leftarrow t+1$}}
        \ENDWHILE
        \ENSURE $\bm{x}_{t}$
    \end{algorithmic}
\end{algorithm}

\section{Simulation Results} \label{sec:result}

\subsection{Simulation Settings}
In this study, we focus on image denoising as the simplest case, assuming that the observation matrix $\bm{A}$ is the identity matrix.
We used 20 monochrome images of size $256\times256$ pixels as original images. 
Since current optical analog circuit development envisions computations in the order of tens to hundreds of dimensions, we divide the images into non-overlapping patches and apply the restoration algorithm to each patch individually in our simulations. 
Specifically, each image was divided into 256 patches, resulting in a patch size of $16\times16$ pixels.

Observed images were generated by adding Gaussian noise with zero mean and a standard deviation of $10/255$ to the original images. 
The initial values for ADMM and PDS were set to $\bm{x}_{0}=\bm{y}$, $\bm{z}_0=\bm{0}$, and $\bm{v}_0=\bm{0}$.
Furthermore, since the typical signal power in optical fiber transmission is approximately 0.001, the additive noise from optical amplifiers was modeled as Gaussian noise with zero mean and a variance scaled to 1000 times the values shown in TABLE~\ref{table(noise)}.

\subsection{Simulation Results}
\begin{figure}[t!]
    \centering
    
    \begin{subfigure}{0.5\textwidth}
        \centering
        \includegraphics[height=4.3cm]{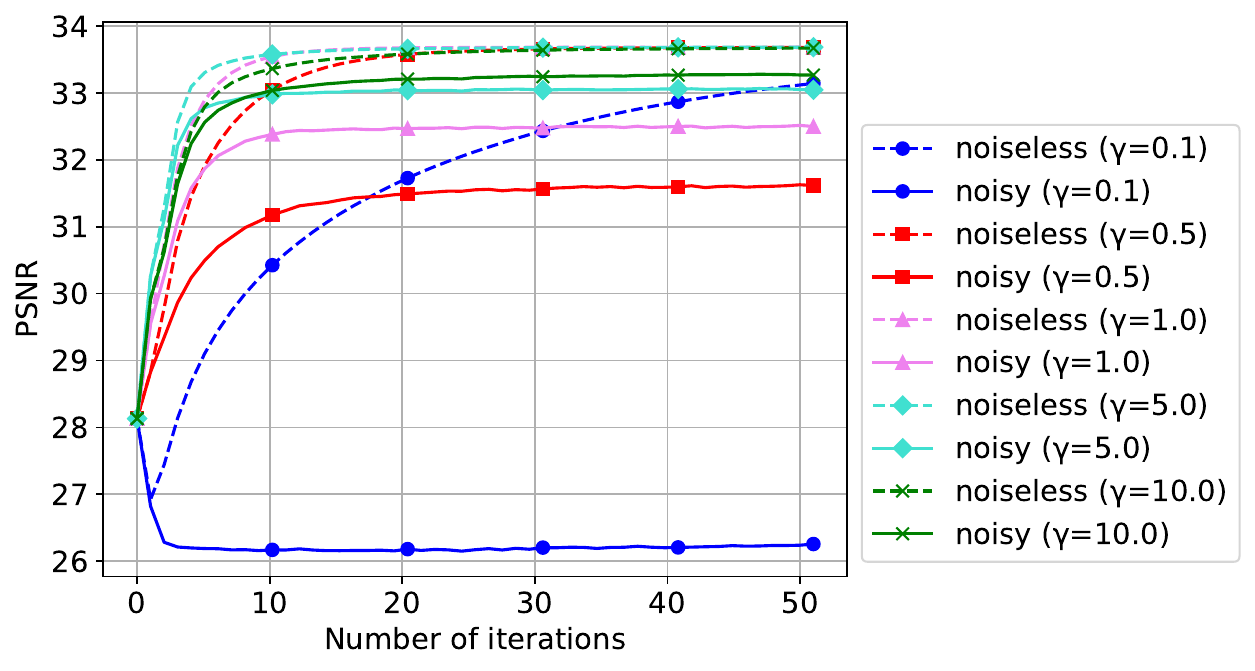}
        \caption{PSNR}
        \label{PSNR-gamma(ADMM)}
    \end{subfigure}
    \begin{subfigure}{0.5\textwidth}
        \centering
        \includegraphics[height=4.3cm]{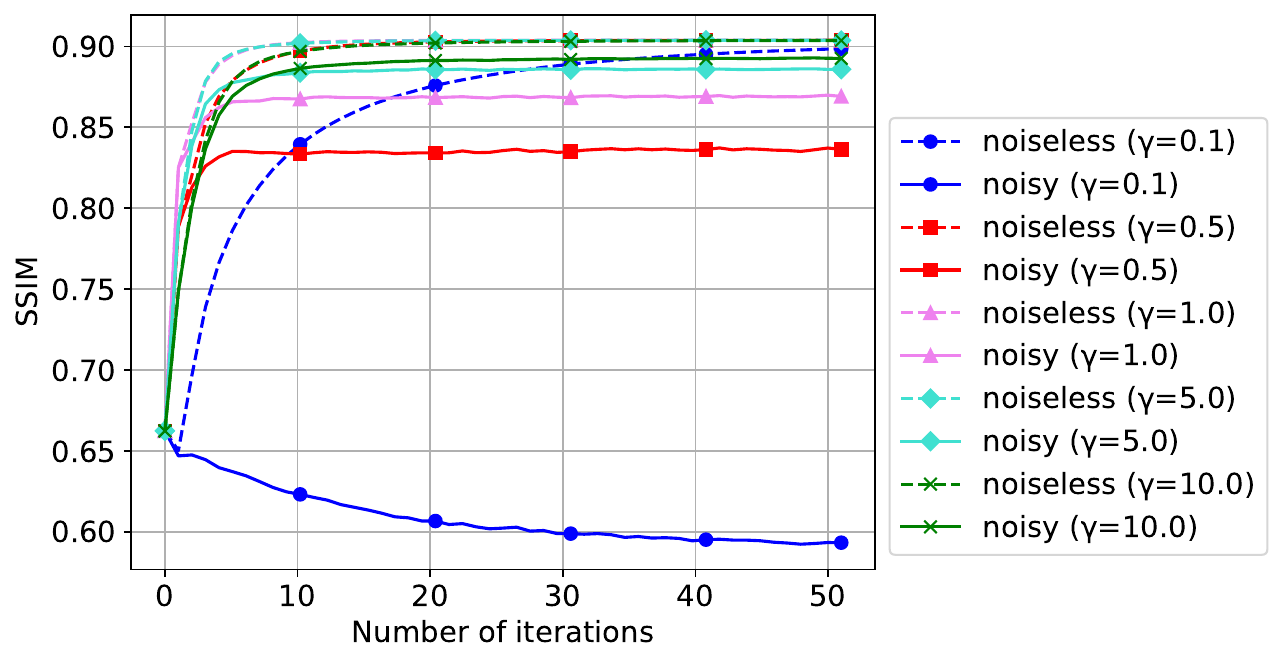}
        \caption{SSIM}
        \label{SSIM-gamma(ADMM)}
    \end{subfigure}
    
    \caption{PSNR and SSIM of ADMM-based algorithm with different values of $\gamma$}
    \label{ADMM-graph}
\end{figure}

Fig.~\ref{ADMM-graph} shows the peak signal-to-noise ratio (PSNR) and structural similarity index measure (SSIM) of ADMM-based algorithm for different values of $\gamma$. 
The regularization parameter is set to $\lambda=0.03$. 
The PSNR and SSIM values in Figs.~\ref{ADMM-graph}(\subref{PSNR-gamma(ADMM)}) and~\ref{ADMM-graph}(\subref{SSIM-gamma(ADMM)}) show the average performance over all divided patches. 
Here, ``noiseless'' refers to the case in which optical amplifier noise is not considered, while ``noisy'' refers to the case in which it is included. 
The impact of circuit noise introduced by multiplication by $1/\gamma$ is also taken into account.
As shown in Fig.~\ref{ADMM-graph}, when the circuit noise is not considered, ADMM converges to almost the same solution with $50$ iterations for $\gamma = 0.1, 0.5, 1.0, 5.0$ and $10.0$. 
In contrast, when the circuit noise is taken into account, the final performance depends on the choice of $\gamma$. 
In the figure, $\gamma = 10$ yields the best results in terms of PSNR and SSIM.
With appropriately selected parameters, the accuracy degradation due to optical amplifier noise is only \SI{0.4}{\deci\bel} in PSNR and 0.008 in SSIM.

\begin{figure}[t!]
    \centering
    
    \begin{subfigure}{0.5\textwidth}
        \centering
        \includegraphics[height=4.3cm]{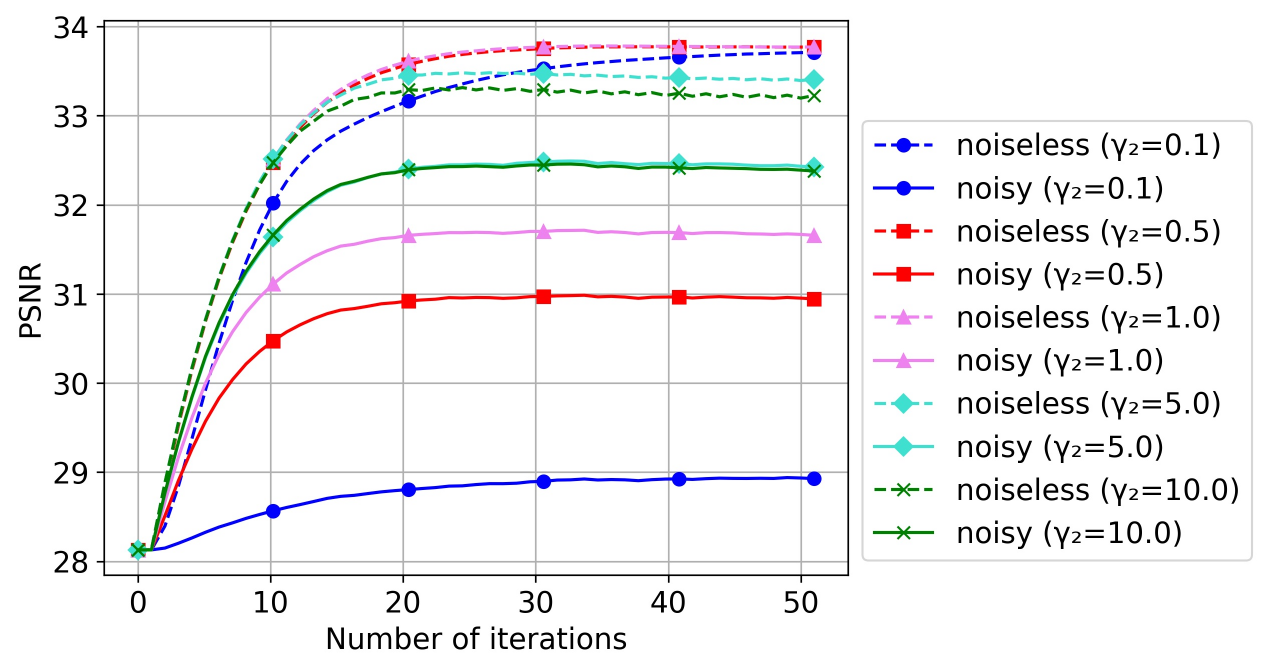}
        \caption{PSNR}
        \label{PSNR-gamma(PDS)}
    \end{subfigure}
    \begin{subfigure}{0.5\textwidth}
        \centering
        \includegraphics[height=4.3cm]{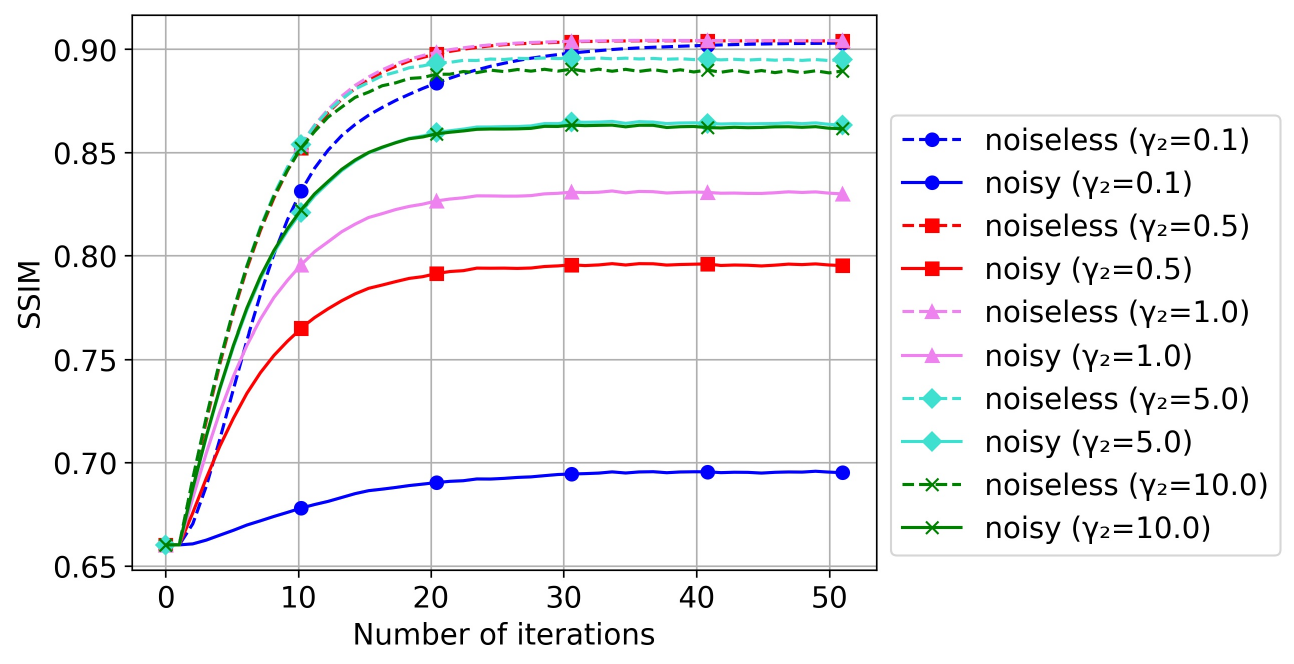}
        \caption{SSIM}
        \label{SSIM-gamma(PDS)}
    \end{subfigure}
    
    \caption{PSNR and SSIM of PDS-based algorithm with different values of $\gamma_{2}$}
    \label{PDS-graph}
\end{figure}

Fig.~\ref{PDS-graph} shows the PSNR and SSIM of PDS for different values of $\gamma_{2}$ (with $\lambda=0.03, \gamma_{1}=0.1$).
As in the case of ADMM, the PSNR and SSIM values show the average performance over all divided patches, and the impact of circuit noise due to multiplication by $\gamma_{1}$ and $\gamma_{2}$ is also taken into account.
As shown in Fig.~\ref{PDS-graph}, when the circuit noise is not considered, PDS with $50$ iterations achieve the best performance when $\gamma_{2}=1.0$. 
On the other hand, when the circuit noise is taken into account, $\gamma_{2}=5.0$ yields the best PSNR and SSIM. 
From the best performance achieved under each setting (i.e., $\gamma_2=5.0$ for the noisy case versus $\gamma_2=1.0$ for the noiseless case), the performance degradation in PSNR is \SI{1.3}{\deci\bel} and the difference in SSIM is 0.032.

\begin{figure}[t!]
    \centering
    
    \begin{subfigure}{0.5\textwidth}
        \centering
        \includegraphics[height=6cm]{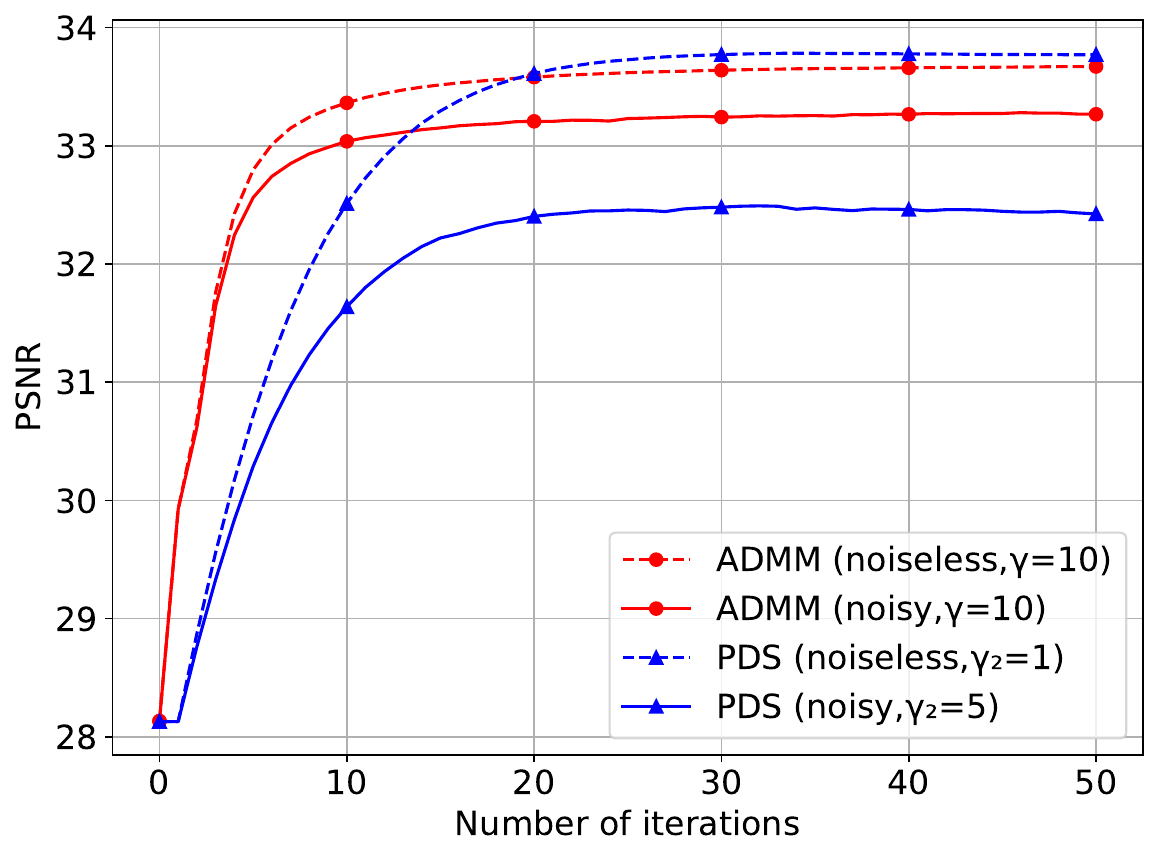}
        \caption{PSNR}
        \label{PSNR-ADMMPDS}
    \end{subfigure}
    \begin{subfigure}{0.5\textwidth}
        \centering
        \includegraphics[height=6cm]{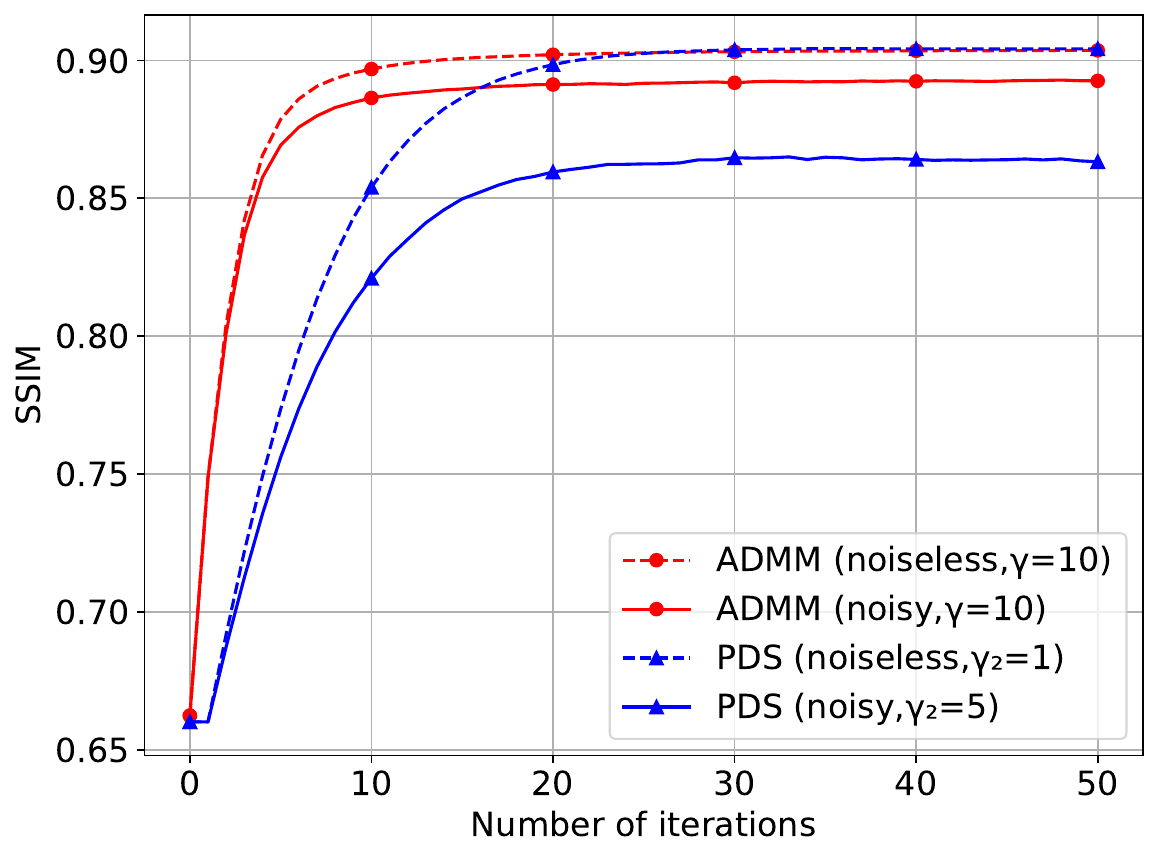}
        \caption{SSIM}
        \label{SSIM-ADMMPDS}
    \end{subfigure}
    
    \caption{Comparison of restoration accuracy between ADMM and PDS.}
    \label{ADMMPDS-graph}
\end{figure}

Fig.~\ref{ADMMPDS-graph} compares the restoration accuracy of ADMM and PDS with their good parameters. 
In the absence of optical amplifier noise, the PSNR and SSIM values for ADMM and PDS converge to almost the same values.
The slight difference in final accuracy is probably because the number of iterations $K=50$ is insufficient.
When optical amplifier noise is taken into account, ADMM achieves higher restoration accuracy than PDS.
This might be because PDS requires a complex optical analog circuit configuration with more optical amplifiers than ADMM, leading to greater noise accumulation.

\begin{figure*}[t!]
    \centering
    \begin{subfigure}[t]{0.16\textwidth}
        \centering
        \includegraphics[width=\textwidth]{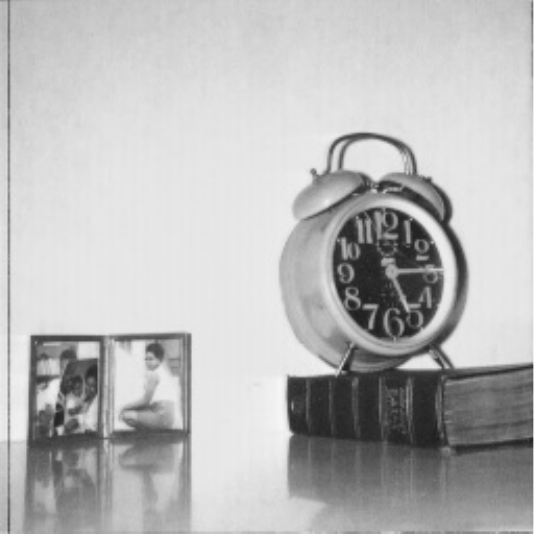}
        \caption{Original image}
    \end{subfigure}
    \hfill
    \begin{subfigure}[t]{0.16\textwidth}
        \centering
        \includegraphics[width=\textwidth]{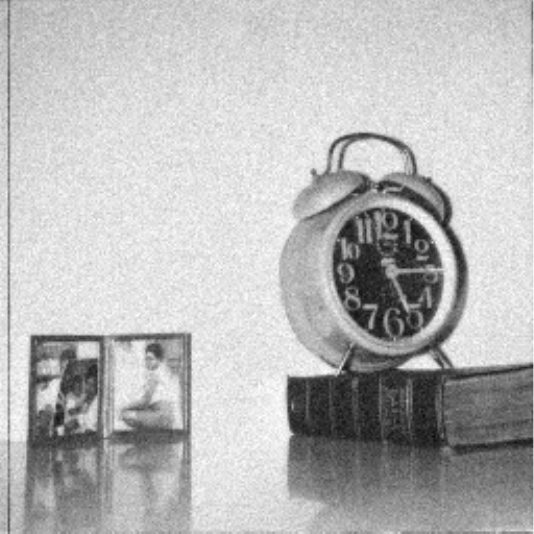}
        \caption{Observed image\\PSNR: 28.13 \si{\deci\bel}\\SSIM: 0.59}
    \end{subfigure}
    \hfill
    \begin{subfigure}[t]{0.16\textwidth}
        \centering
        \includegraphics[width=\textwidth]{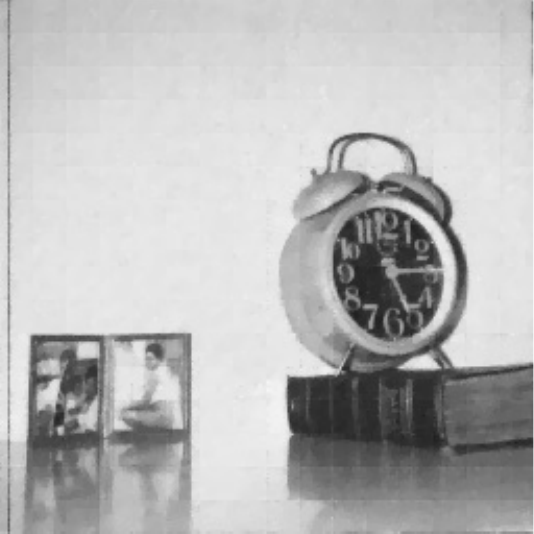}
        \caption{ADMM (noiseless)\\PSNR: 33.39 \si{\deci\bel}\\SSIM: 0.92}
    \end{subfigure}
    \hfill
    \begin{subfigure}[t]{0.16\textwidth}
        \centering
        \includegraphics[width=\textwidth]{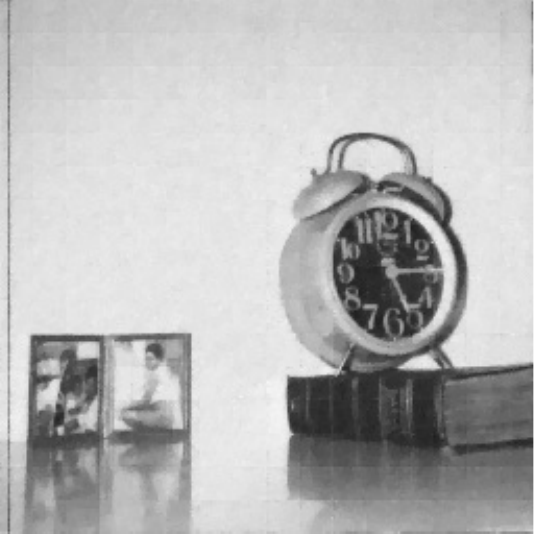}
        \caption{ADMM (noisy)\\PSNR: 33.17 \si{\deci\bel}\\SSIM: 0.91}
    \end{subfigure}
    \hfill
    \begin{subfigure}[t]{0.16\textwidth}
        \centering
        \includegraphics[width=\textwidth]{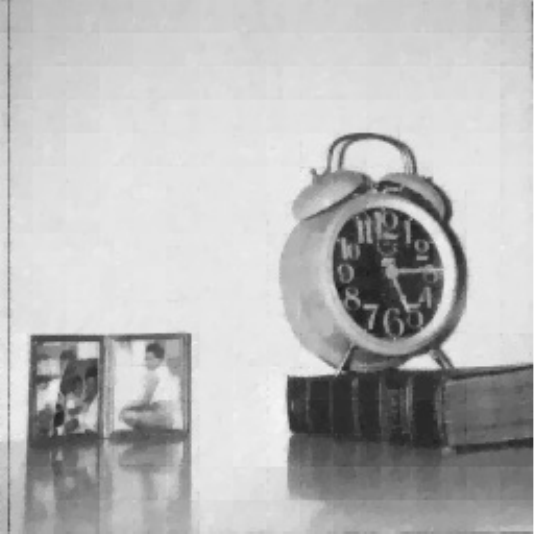}
        \caption{PDS (noiseless)\\PSNR: 33.46 \si{\deci\bel}\\SSIM: 0.93}
    \end{subfigure}
    \hfill
    \begin{subfigure}[t]{0.16\textwidth}
        \centering
        \includegraphics[width=\textwidth]{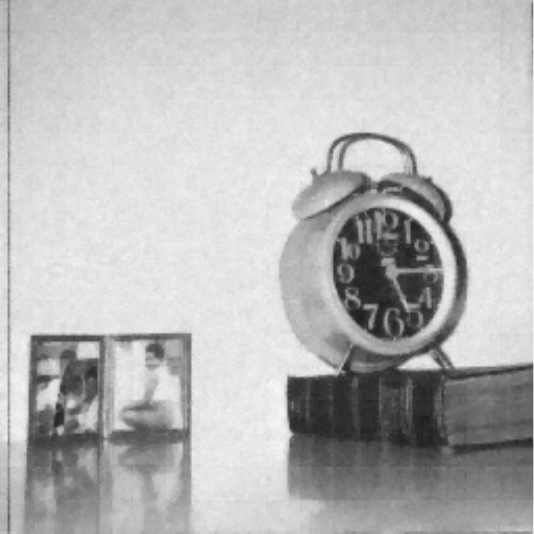}
        \caption{PDS (noisy)\\PSNR: 32.49 \si{\deci\bel}\\SSIM: 0.86}
    \end{subfigure}
    \caption{Original, observed, and restored images (noiseless / noisy) for `Clock' image}
    \label{image-clock}
\end{figure*}

\begin{figure*}[t!]
    \centering
    \begin{subfigure}[t]{0.16\textwidth}
        \centering
        \includegraphics[width=\textwidth]{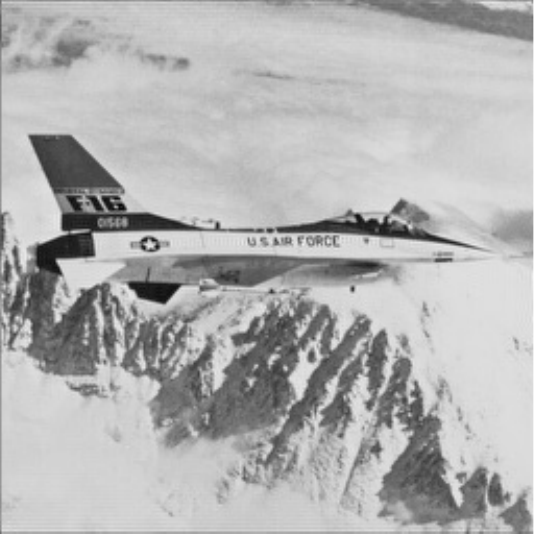}
        \caption{Original image}
    \end{subfigure}
    \hfill
    \begin{subfigure}[t]{0.16\textwidth}
        \centering
        \includegraphics[width=\textwidth]{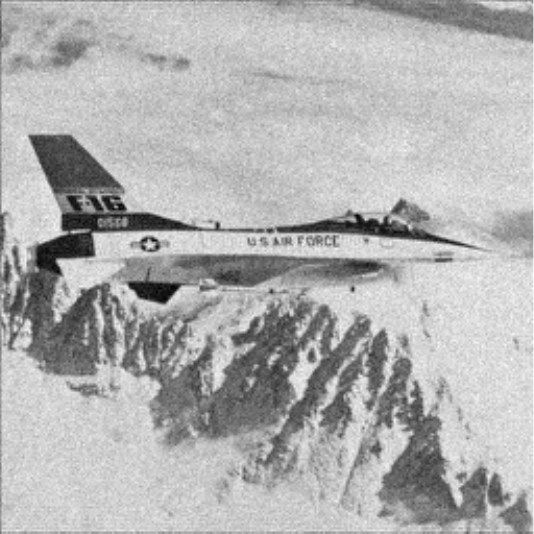}
        \caption{Observed image\\PSNR: 28.12 \si{\deci\bel}\\SSIM: 0.68}
    \end{subfigure}
    \hfill
    \begin{subfigure}[t]{0.16\textwidth}
        \centering
        \includegraphics[width=\textwidth]{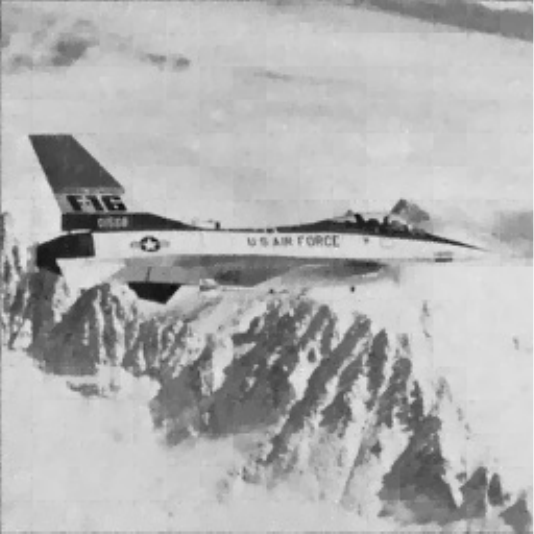}
        \caption{ADMM (noiseless)\\PSNR: 31.48 \si{\deci\bel}\\SSIM: 0.90}
    \end{subfigure}
    \hfill
    \begin{subfigure}[t]{0.16\textwidth}
        \centering
        \includegraphics[width=\textwidth]{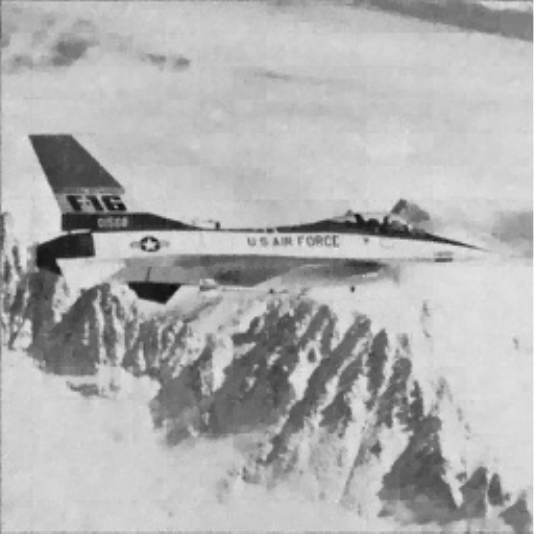}
        \caption{ADMM (noisy)\\PSNR: 31.38 \si{\deci\bel}\\SSIM: 0.89}
    \end{subfigure}
    \hfill
    \begin{subfigure}[t]{0.16\textwidth}
        \centering
        \includegraphics[width=\textwidth]{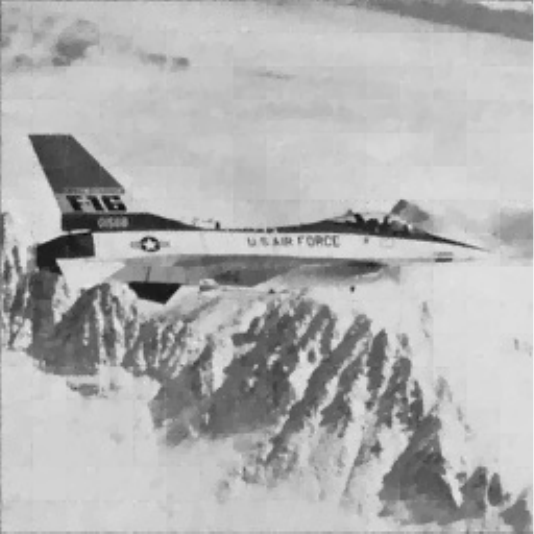}
        \caption{PDS (noiseless)\\PSNR: 31.52 \si{\deci\bel}\\SSIM: 0.90}
    \end{subfigure}
    \hfill
    \begin{subfigure}[t]{0.16\textwidth}
        \centering
        \includegraphics[width=\textwidth]{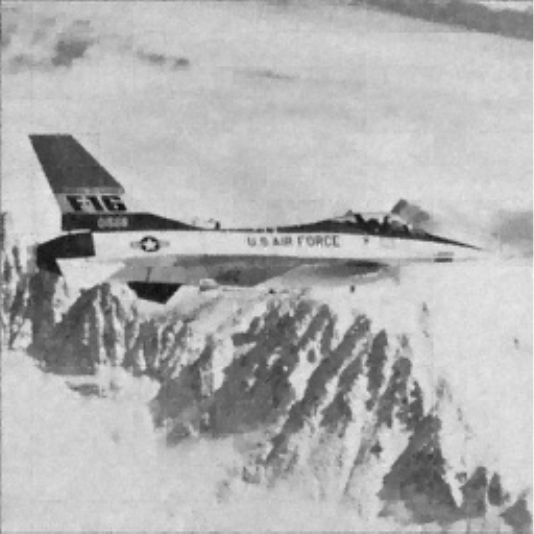}
        \caption{PDS (noisy)\\PSNR: 30.93 \si{\deci\bel}\\SSIM: 0.86}
    \end{subfigure}
    \caption{Original, observed, and restored images (noiseless / noisy) for `Airplane' image}
    \label{image-airplane}
\end{figure*}

\begin{figure*}[t!]
    \centering
    \begin{subfigure}[t]{0.16\textwidth}
        \centering
        \includegraphics[width=\textwidth]{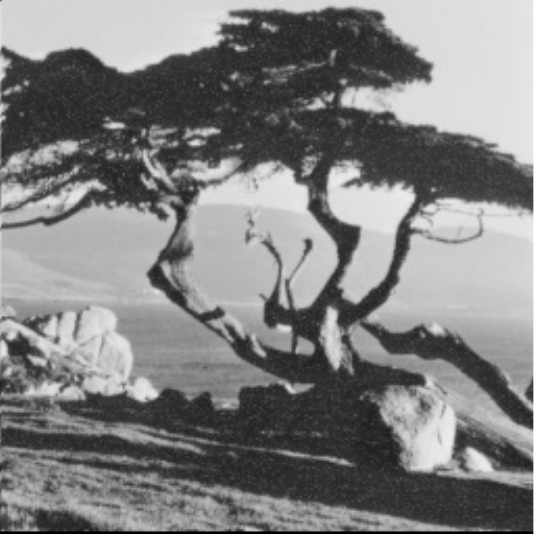}
        \caption{Original image}
    \end{subfigure}
    \hfill
    \begin{subfigure}[t]{0.16\textwidth}
        \centering
        \includegraphics[width=\textwidth]{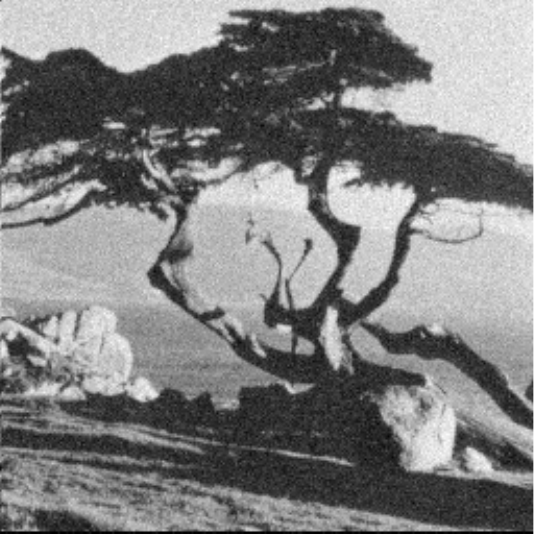}
        \caption{Observed image\\PSNR: 28.13 \si{\deci\bel}\\SSIM: 0.76}
    \end{subfigure}
    \hfill
    \begin{subfigure}[t]{0.16\textwidth}
        \centering
        \includegraphics[width=\textwidth]{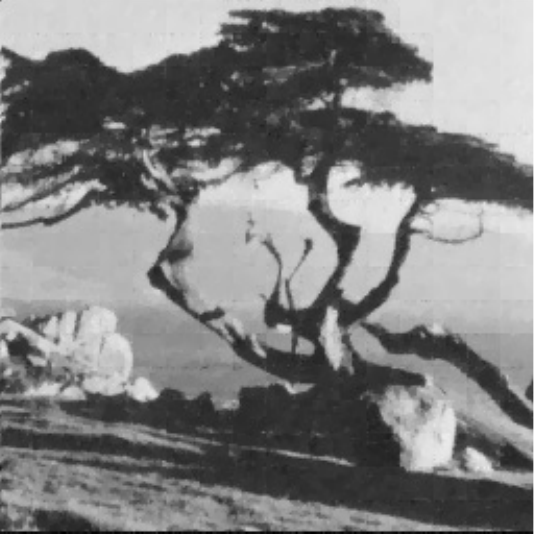}
        \caption{ADMM (noiseless)\\PSNR: 30.65 \si{\deci\bel}\\SSIM: 0.89}
    \end{subfigure}
    \hfill
    \begin{subfigure}[t]{0.16\textwidth}
        \centering
        \includegraphics[width=\textwidth]{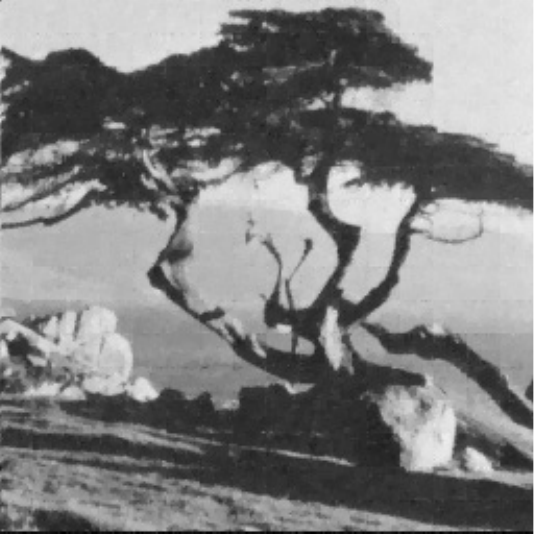}
        \caption{ADMM (noisy)\\PSNR: 30.59\si{\deci\bel}\\SSIM: 0.88}
    \end{subfigure}
    \hfill
    \begin{subfigure}[t]{0.16\textwidth}
        \centering
        \includegraphics[width=\textwidth]{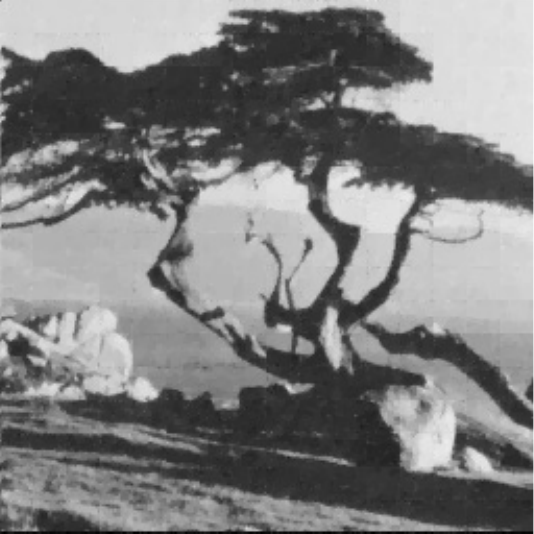}
        \caption{PDS (noiseless)\\PSNR: 30.68 \si{\deci\bel}\\SSIM: 0.89}
    \end{subfigure}
    \hfill
    \begin{subfigure}[t]{0.16\textwidth}
        \centering
        \includegraphics[width=\textwidth]{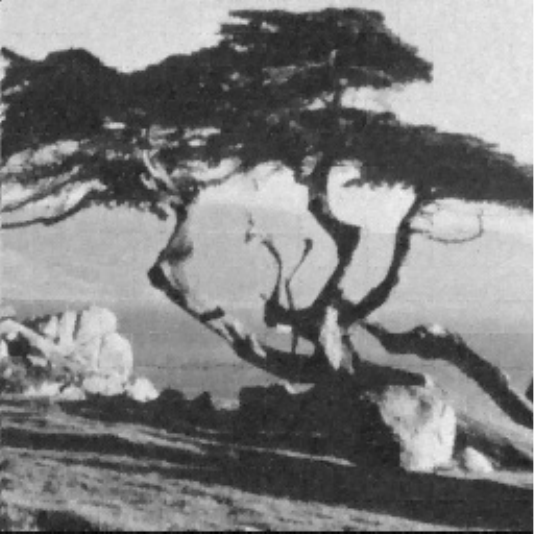}
        \caption{PDS (noisy)\\PSNR: 30.29 \si{\deci\bel}\\SSIM: 0.87}
    \end{subfigure}
    \caption{Original, observed, and restored images (noiseless / noisy) for `Tree' image}
    \label{image-tree}
\end{figure*}

\begin{figure*}[t!]
    \centering
    \begin{subfigure}[t]{0.16\textwidth}
        \centering
        \includegraphics[width=\textwidth]{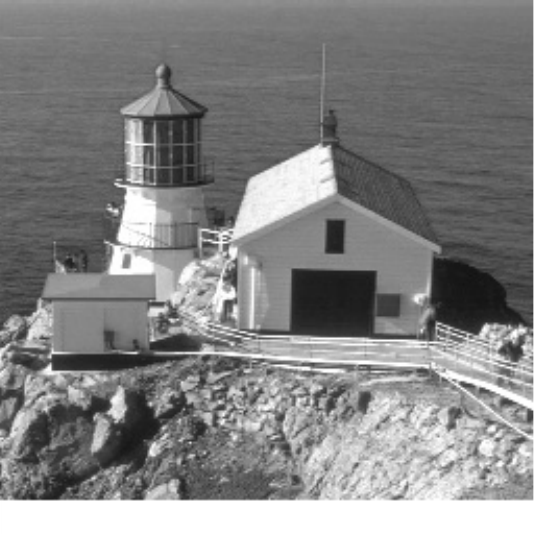}
        \caption{Original Image}
    \end{subfigure}
    \hfill
    \begin{subfigure}[t]{0.16\textwidth}
        \centering
        \includegraphics[width=\textwidth]{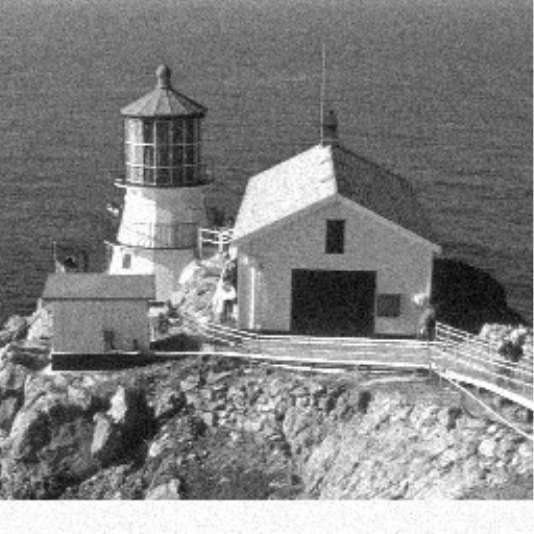}
        \caption{Observed Image\\PSNR: 28.30 \si{\deci\bel}\\SSIM: 0.75}
    \end{subfigure}
    \hfill
    \begin{subfigure}[t]{0.16\textwidth}
        \centering
        \includegraphics[width=\textwidth]{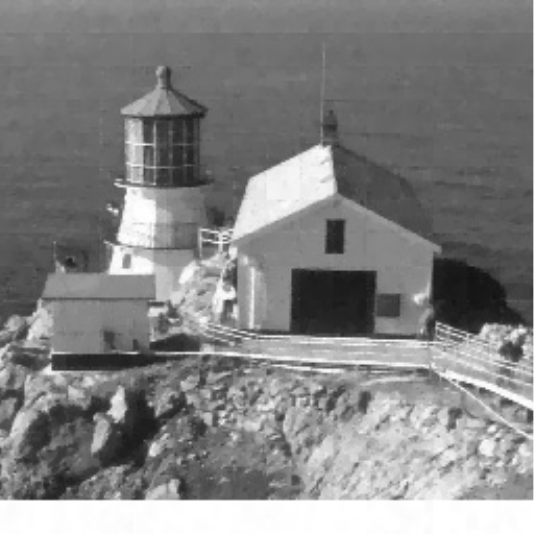}
        \caption{ADMM (noiseless)\\PSNR: 29.82 \si{\deci\bel}\\SSIM: 0.88}
    \end{subfigure}
    \hfill
    \begin{subfigure}[t]{0.16\textwidth}
        \centering
        \includegraphics[width=\textwidth]{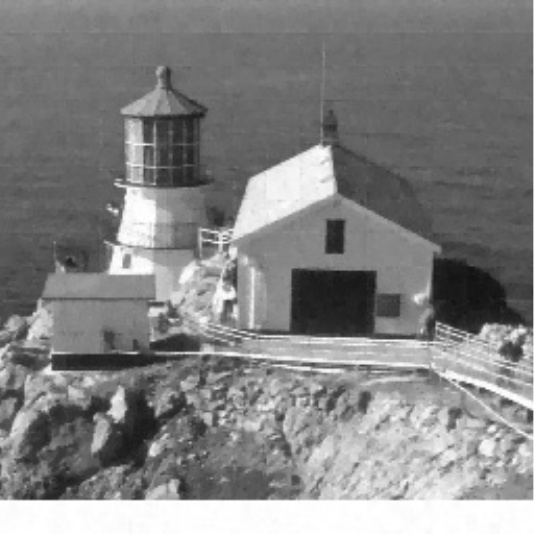}
        \caption{ADMM (noisy)\\PSNR: 29.78\si{\deci\bel}\\SSIM: 0.87}
    \end{subfigure}
    \hfill
    \begin{subfigure}[t]{0.16\textwidth}
        \centering
        \includegraphics[width=\textwidth]{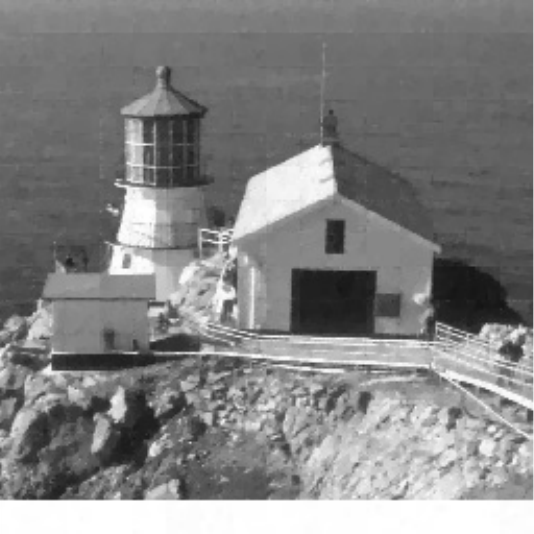}
        \caption{PDS (noiseless)\\PSNR: 29.90 \si{\deci\bel}\\SSIM: 0.88}
    \end{subfigure}
    \hfill
    \begin{subfigure}[t]{0.16\textwidth}
        \centering
        \includegraphics[width=\textwidth]{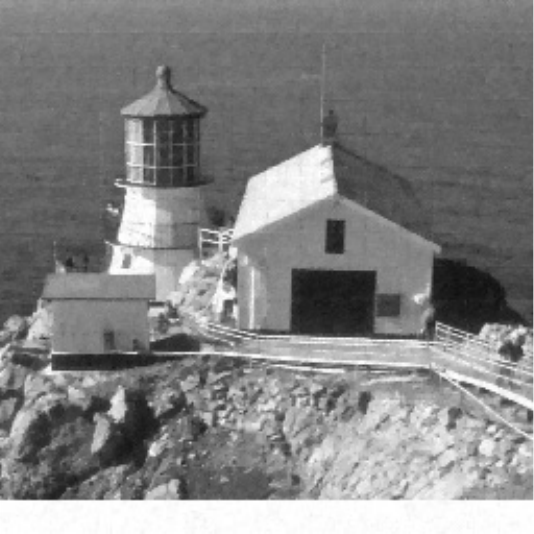}
        \caption{PDS (noisy)\\PSNR: 29.55 \si{\deci\bel}\\SSIM: 0.86}
    \end{subfigure}
    \caption{Original, observed, and restored images (noiseless / noisy) for `Lighthouse' image}
    \label{image-lighthouse}
\end{figure*}

Figures~\ref{image-clock}--\ref{image-lighthouse} show the original images, observed images, restored images using ADMM ($\gamma=10, \lambda=0.03$) and PDS. 
For the PDS results, the parameters were set to $(\gamma_{1}=0.1, \gamma_{2}=1, \lambda=0.03)$ for the noiseless case and $(\gamma_{1}=0.1, \gamma_{2}=5, \lambda=0.03)$ for the noisy case. 
From these figures and the PSNR and SSIM values, we can see that the noise in the observed images can be reduced even when circuit noise is considered.
\section{Conclusion} \label{sec:summary}
In this study, we investigated two algorithms based on total variation regularization for image restoration in optical analog circuits.
Considering the noise added by optical amplifiers at each iteration of the algorithms, we compared the restoration accuracy with and without additive circuit noise through computer simulations.
Simulation results show that the noisy images can be effectively denoised even when the circuit noise introduced by optical amplifiers is taken into account.
Furthermore, our experiments indicated that ADMM achieved higher restoration accuracy compared to PDS in the noisy case.

Future work includes extending the proposed approach to image restoration problems beyond denoising and the investigating implementation methods for proximal operator computations in electrical circuits. 
Additionaly, analyzing the impact of inaccuracies in matrix-vector products is also important direction for future development.

%
%------------------
% References
%------------------
\bibliographystyle{IEEEtran}
\bibliography{reference}

\vfill

\end{document}